# Quantum Test of the Equivalence Principle and Space-Time aboard the International Space Station

**Jason Williams, Sheng-wey Chiow, and Nan Yu**
*Jet Propulsion Laboratory, California Institute of Technology, Pasadena, CA 91109, USA*
**Holger Müller**
*Department of Physics, University of California, Berkeley, CA 94720, USA*
E-mail: Nan.Yu@jpl.nasa.gov and hm@berkeley.edu

**Abstract**
We describe the Quantum Test of the Equivalence principle and Space Time (QTEST), a concept for an atom interferometry mission on the International Space Station (ISS). The primary science objective of the mission is a test of Einstein's equivalence principle with two rubidium isotope gases at a precision of better than $10^{-15}$, a 100-fold improvement over the current limit on equivalence principle violations, and over 1,000,000 fold improvement over similar quantum experiments demonstrated in laboratories. Distinct from the classical tests is the use of quantum wave packets and their expected large spatial separation in the QTEST experiment. This dual species atom interferometer experiment will also be sensitive to time-dependent equivalence principle violations that would be signatures for ultralight dark-matter particles. In addition, QTEST will be able to perform photon recoil measurements to better than $10^{-11}$ precision. This improves upon terrestrial experiments by a factor of 100, enabling an accurate test of the standard model of particle physics and contributing to mass measurement, in the proposed new international system of units (SI), with significantly improved precision. The predicted high measurement precision of QTEST comes from the microgravity environment on ISS, offering extended free fall times in a well-controlled environment. QTEST plans to use high-flux, dual-species atom sources, and advanced cooling schemes, for $N > 10^6$ non-condensed atoms of each species at temperatures below 1nK. Suppression of systematic errors by use of symmetric interferometer configurations and rejection of common-mode errors drives the QTEST design. It uses Bragg interferometry with a single laser beam at the "magic" wavelength, where the two isotopes have the same polarizability, for mitigating sensitivities to vibrations and laser noise, imaging detection for correcting cloud initial conditions and maintaining contrast, modulation of the atomic hyperfine states for reduced sensitivity to magnetic field gradients, two source-regions for simultaneous time reversal measurements and redundancy, and modulation of the gravity vector using a rotating platform to reduce otherwise difficult systematics to below $10^{-16}$.

# 1 Introduction

Gravity is the least-tested force of nature. For example, even the best solar system tests (*e.g.*, the Shapiro time delay to Cassini) reach $10^{-5}$ sensitivity in the parameterized post-Newtonian (PPN) formalism [1], while the theory of quantum electrodynamics has some of its predictions confirmed at the part-per-billion level. Further, some predicted effects, such as gravitational waves, are yet to be observed directly. The lack of a quantum theory of gravity and the unexplained origins of dark matter and dark energy hint that the current understanding of gravity may not be complete. Testing the equivalence principle (EP) is one of the key methods for probing gravity physics. Without the EP, general relativity cannot be correct. However, the EP is (or may be) violated in many theories that attempt to join gravity with the standard model of particle physics (*e.g.*, string theory, loop quantum gravity, higher dimensions, brane worlds) [2–5]. Tests of the EP are also sensitive to new physics, *e.g.*, dilatons or moduli [6]. Such tests may thus be one of the best chances to detect new physics beyond the standard model.

Recent advances in ultra-cold atom physics and atom interferometry have provided new measurement capabilities [7–13], with which the EP can be tested to unprecedented precisions. Already, laboratory experimental investigations have been carried out with atom interferometers for tests of the equivalence principle [14–17], for precise photon recoil measurements [18,19], and for tests of inverse square laws of gravity [20–22]. Quantum Test of the Equivalence principle and Space Time (QTEST) will fundamentally be a set of light-pulse atom interferometer (AI) experiments. Its primary science objective is to measure the differential gravitational acceleration between $^{87}$Rb and $^{85}$Rb to a precision better than one part in $10^{15}$. Differentiating from classical bulk matter EP tests, AI measurements in QTEST allow access to rare isotopes, probe spin-gravity coupling, study coherent quantum wave packets, and explore the influence of gravity in quantum mechanics [23]. A secondary science objective is to improve photon recoil measurements. This can be accomplished by simply changing the interferometer measurement sequences. The improvement in the photon recoil measurement will result in the most stringent test of quantum electrodynamics by verifying the electron's gyromagnetic ratio, the most precise prediction made in any science [24], and can lead to the most precise realization of the unit of mass in the revised international system of units [25].

Space microgravity environments provide a nearly ideal platform for atom interferometry. Indeed, much of the precision gain of QTEST over the ground experiments comes from the operation environment on ISS. The most recognized benefit is the extended free fall duration, allowing long interrogation times. Since, for Mach-Zehnder AI, the inertial-measurement sensitivity scales quadratically with the interrogation time, the sensitivity gain is significant. Just as important for QTEST is the ability to modulate gravity by rotating the setup on a rotation stage, which helps strongly reduce the systematics by phase sensitive detection. The lack of a large, static gravitational force in space also minimizes the necessary length of the atom trajectories and, therefore, simplifies and enhances environmental isolation and control. This aspect can be most appreciated by the fact that, to achieve an equivalent of 10 s total interrogation time on the ground would need over 100 m of apparatus size [13], whereas the same sensitivity in microgravity can be realized in a sensor measuring less than 1 m, planned for QTEST. Furthermore, a space environment will provide opportunities for new science. The high velocity of the space station makes it possible to perform precision tests of the relativistic properties of gravity [26,27], i.e., terms in test theories that enter the signal proportional to the gravitational acceleration and a power of the velocity with respect to the source mass of the gravitational potential. Finally, the long interrogation time in a stable environment makes QTEST optimally sensitive to time-dependent EP violating terms that will narrow the range of feasible dark matter candidates [28].

QTEST strives to maximize common-mode suppression of systematic errors by using a highly symmetric measurement approach. This includes the use of two Rb isotopes with a single Bragg laser beam addressing both isotopes simultaneously. This helps to suppress the sensitivity to vibrational noise, which is necessary on the ISS where the vibration environment is not the most quiet. It will also use two dual-species sources, enabling two dual-species atom interferometers (four AIs in total) to be interrogated simultaneously in a time-reversed configuration. This reduces systematic sources that are sensitive to the photon recoil and to the AI laser orientations. Together with a rotating platform based gravity modulation, these are the key design points that enable QTEST to achieve its primary science objective for the EP test.

This paper is organized as follows: Section 2 reviews the major science motivations. Section 3 presents an overview of the QTEST mission concept. Section 4 details the experimental implementation with emphasis on the approaches and concepts that are unique to QTEST. Section 5 discusses statistical and systematic error sources for the EP test and their mitigations.

## 2  Science Motivations

### 2.1  Test of the equivalence principle

While everybody can feel the force of gravity, it is a stranger among the four forces known in physics. For one, it accelerates all objects in the same way, regardless of, *e.g.*, their structure, mass, charge, or composition. This observation, known as the weak equivalence principle (WEP) or universality of free fall (UFF), has been extended to the Einstein equivalence principle, which states that gravity is locally equivalent to the effects of being in an accelerated frame of reference for any experiment. "Equivalence" here means that there is no experiment that can tell the two apart and "locally" means that the statement is restricted to sufficiently small regions of spacetime, such that gravity gradients are negligible.

The validity of the equivalence principle is deeply ingrained in the current gravitational theory; without it, general relativity could not be true [29]. Experimental tests of Lorentz invariance [30], local position invariance [31], and the weak equivalence principle [32] have shown that nature adheres closely to this principle. But it doesn't have to be so. Modern theories to unify gravity and quantum mechanics operate at energy scales so high that direct experiments are impossible. Violations of the equivalence principle, which can be parameterized by the standard model extension (SME) [33–35], are among the most promising candidates for probing Planck-scale physics at low energy [5,36–38]. For a review of the science motivations of next-generation, ultra-high precision equivalence principle tests and their expected implications for the foundations of contemporary gravitational theories, refer to [2,6,23] and references therein.

WEP violations are quantified by the Eötvös parameter, the normalized differential acceleration of two test particles (A and B) under the influence of gravity:

$$\eta(A,B) = 2\frac{g^A - g^B}{g^A + g^B} \tag{1}$$

Following the formalism in Reference [39], limits on the minimal gravitational SME coefficients $\alpha(\bar{a}_{\text{eff}}^w)_0$ and $(\bar{c}^w)_{00}$, which describe the size of various types of Lorentz violations in the test framework of the SME, can be extrapolated from $\eta$ by the definition of the total gravitational acceleration of each test mass ($m^T$):

$$g^T = g(1-\beta^T), \qquad \beta^T = \frac{2\alpha}{m^T}\sum_w N^w (\bar{a}_{\text{eff}}^w)_0 - \frac{2}{3m^T}\sum_w N^w m^w (\bar{c}^w)_{00} \tag{2}$$

| | | | |
|---|---|---|---|
| $\alpha(\bar{a}_{\text{eff}}^{p+e})_0$ | $5 \times 10^{-14}$ GeV | $(\bar{c}^p)_{00}$ | $1.5 \times 10^{-13}$ GeV |
| $\alpha(\bar{a}_{\text{eff}}^{n})_0$ | $5 \times 10^{-14}$ GeV | $(\bar{c}^n)_{00}$ | $1.5 \times 10^{-13}$ GeV |
| $\alpha(\bar{a}_{\text{eff}}^{p+e-n})_X$ | $\{10^{-9}$ GeV$\}$ | $(\bar{c}^e)_{00}$ | $3 \times 10^{-10}$ GeV |
| $\alpha(\bar{a}_{\text{eff}}^{p+e-n})_{Y+Z}$ | $\{10^{-9}$ GeV$\}$ | $(\bar{c}^n)_Q$ | $\{10^{-13}$ GeV$\}$ |
| $\alpha(\bar{a}_{\text{eff}}^{p+e-n})_Y$ | $\{10^{-7}$ GeV$\}$ | $(\bar{c}^n)_{TJ}$ | $\{10^{-9}$ GeV$\}$ |
| $\alpha(\bar{a}_{\text{eff}}^{p+e-n})_Z$ | $\{10^{-7}$ GeV$\}$ | | |

**Table 1**: SME coefficient sensitivities, extrapolated from the projected QTEST measurement sensitivity of 10$^{-15}$. The numerical coefficients can be refined by taking into account the nuclear binding energy [40]. Values in braces represent order-of-magnitude estimates for sensitivities arising due to the orbital motion of the ISS. See [41], section VIII, for details and definitions of coordinates.

Here, the superscript *w* takes the values *e, p,* and *n* representing the electron, proton, and neutron respectively. For neutral matter, with equal electron and proton numbers ($N^e = N^p$), the notation can be simplified by using $(\bar{a}_{\text{eff}}^{p+e})_0 = (\bar{a}_{\text{eff}}^{p})_0 + (\bar{a}_{\text{eff}}^{e})_0$. The projected sensitivity limits for these SME coefficients in QTEST are given in Table 1.

Measurements with space-based platforms also open the door for unique and/or enhanced searches for EP violations, parameterized by $\alpha(\bar{a}_{\text{eff}}^w)_\mu$ and $(\bar{c}^w)_{\mu\nu}$ beyond the zero components ($\mu = \nu = 0$). In addition to the long free fall times available in space, the orientation and rotation of the test-bodies can be decoupled from the direction of gravity, useful for detection of several Lorentz-violating possibilities. Extrapolating from theoretical analyses by V. Kostelecký and J. Tasson [41], it is clear that the projected sensitivity of QTEST could also improve over the current limits of non-zero SME components [42] by orders of magnitude (see Table 1).

High sensitivity AI sensors, in orbit around the earth, will also help to detect nonstandard signals for deviations from general relativity. For example, the EP violations may depend upon the test particles' motion relative to the source of the gravitational field (here, the earth or the sun). In technical language, such effects are encoded in higher-mass dimension operators of the SME [43]. These effects have never been the subjects of an experimental study, though incomplete bounds can be derived from solar system observations like perihelion precession [44]. A scenario in which such terms may be nonzero is given by theories that attempt to explain cosmic acceleration by a scalar field (Chameleons, *f(R)*, Gallileons,…), which could lead to order-one violation of the EP for extended objects such as galaxies or constituents thereof but not in terrestrial experiments [45]. Such theories evade detection by the chameleon mechanism, which suppresses any effects in environments of high density. This renders them invisible in classical experiments but not in quantum experiments [46]. Atom interferometers in orbit, such as QTEST, have the additional advantage that atoms can remain close to fixed objects that generate the chameleon force for seconds instead of tens of milliseconds, producing a sensitivity gain of ten thousand. In addition, there are other theories such as galilleons, for which no method of detection is currently known. Since effects of these theories are strongly suppressed anywhere near the Earth's or other celestial objects, any detection would have to happen in space [47,48]. This subject is currently under investigation.

*2.1.1 Significance of quantum tests*

By a quantum test, we mean an experiment in which the wavelike properties of matter are used to measure the influence of gravity, *i.e.,* matter-wave interferometers. Such tests are important for technical and fundamental reasons. Among the technical reasons, they allow for unique test mass compositions, such as alkali metals, which cannot be used in classical tests. This broadens the coverage of species for which the UFF can be verified, helping to constrain all modes of equivalence principles that are possible, according to models such as the SME [40]. Furthermore, like in atomic clocks, the quantum tests allow use of the well-understood and controlled properties of atoms to mitigate systematics. All atoms of one species are alike and are exactly electrically neutral. Further, their interactions with the environment (magnetic and electric fields, radiation, etc.) are well understood and can be controlled to high precision. Neutral atoms are a close approximation to the textbook concept of a light, isolated, non-interacting point mass.

One fundamental difference of quantum tests, as compared to classical ones, is that they are sensitive to the interference of matter waves as they evolve over time. In that way, they go beyond testing the WEP, which is a statement about a particle's center-of-mass motion only. Quantum-based measurements test how gravity enters quantum mechanics, whether the standard treatment of adding the gravitational potential to the Schrödinger equation is an accurate description of nature. The Schrödinger description of gravity is equivalent to a path integral with the action given by the proper time of a particle. For this reason, quantum tests of the EP have been described as a test of the gravitational redshift with matter waves [16,49,50]. Schiff's conjecture [51] states that the WEP implies local position invariance and thus the full EP. If so, the phase of a wave packet is determined by the action along the classical path [52] and quantum tests are sensitive to the same physics as classical ones [53]. After 50 years, however, the conjecture has neither been proven nor disproven, though specialized counterexamples exist [54–56].

Another fundamental difference to classical experiments is the ability to probe spin-dependent gravitational couplings. These have long been studied in the context of theories of gravity with nonvanishing torsion [57]. In the SME [30], such effects are expected to result from the $b_\mu$, $d_{\mu\nu}$, $f_\mu$, $g_{\lambda\mu\nu}$, and $H_{\mu\nu}$ coefficients, which also describe how gravity might couple differently to particles exhibiting different spin-orbit couplings, *i.e.,* having correlated external and spin degrees of freedom. These signals are the result of interactions of the quantum properties of matter with gravity. QTEST could perform the first and/or most sensitive measurement on several of these coefficients.

*2.1.2 Ultralight dark matter and dark energy*

A variety of cosmological and astrophysical measurements have established that dark matter may be a new particle [58]. Weakly Interacting Massive Particles (WIMPs) are popular dark matter candidates that are presently searched for in a variety of experiments [59], but they are not the only possibility. Ultra-light bosons are expected in many frameworks such as string theory [60,61] that attempt to resolve the theoretical issues raised by the standard model. If these particles exist, as a consequence of the misalignment mechanism [62–65], they can easily be the dark matter of the universe but would be hard to detect in conventional WIMP searches. They may, however, give rise to signals that can be detected by QTEST, such as time-dependent equivalence principle violations.

Observational constraints imply that these kinds of dark matter candidates have to be more massive than $mc^2/h=10^{-7}$ Hz [66]; there is no firm upper bound on this mass. It is thus important to search for these particles wherever feasible. However, at large number density, they can be described as classical fields and give rise to new ways of detection. Much like how gravitational waves are detected through their coherent effects on matter instead of the scattering of single gravitons, these dark matter fields also give rise to new coherent effects. Several experiments that leverage the classical nature of these fields have recently been proposed [67–69]. These include searches for axions, axion-like particles and hidden photons [67–69].

In addition to these particles, there is just one other kind of ultra-light boson that can interact with the standard model and still be naturally light. This is the massive B-L gauge boson. It can be described by a field which behaves very similar to an electric field and that we will call the "B-L electric field," in absence of a better terminology. In particular, particles in such a B-L field feel a force proportional to the field strength and a charge. Protons and neutrons have charge +1 while electrons and neutrinos have charge -1. Thus, neutral atoms have a non-zero "B-L charge" equal to their neutron number. If the B-L gauge boson existed and was light, there are stringent bounds on its coupling strength $g$ since this boson can mediate new short distance forces [70].

If the B-L gauge boson was the dark matter, there would be a time varying B-L electric field everywhere in the galaxy, similar to the time varying electromagnetic field associated with the cosmic microwave background. The B-L electric field, while random over long distances and times, is spatially coherent over distances given by the de Broglie wavelength $h/(mv)$ and temporally coherent over time scales $\sim h/(mv^2)$ where $m$ is the mass of the B-L gauge boson and $v/c \sim 10^{-3}$ is the virial velocity of dark matter in the galaxy [67–69], leading to coherence times as long as a week for masses $m \sim 1$ Hz. These coherence times help experiments like QTEST gain sensitivity through integration.

The B-L electric field would then exert a time-varying force on neutral atoms that have a non-zero neutron number. At frequencies $\sim 1$ Hz, to go beyond present bounds, the accelerations can be as large as $10^{-11}$ m/s$^2$, well within the range of atomic accelerometers [28]. One way to search for this kind of dark matter would be to use inertial accelerometer setups such as those used for gravitational wave detection and tests of the EP [28]. Since the force exerted by the dark matter on the atoms depends upon the neutron number, this force will violate the EP. Moreover, since the dark matter field oscillates in time, the EP violation will also oscillate in time. This signal oscillates at a frequency set by fundamental physics, i.e. the dark matter mass. Additionally, the oscillations will remain coherent for about $\sim 1/v^2 \sim 10^6$ periods. This should be helpful in combatting systematics [28].

The WIMP paradigm is currently under stress from null results at the Large Hadron Collider and a variety of dark matter direct detection experiments [71,72]. Since we do not know much about the origin of dark matter, it is important to search for them in the widest possible range. B-L gauge boson dark matter is one of the few other well-motivated dark matter candidates and the methods proposed here is the only way currently known to search for them beyond present experimental bounds.

### 2.2 Photon recoil measurements with atom interferometers

By reprogramming from Mach-Zehnder to Ramsey-Bordé configuration, i.e., firing a π/2- π/2- π/2- π/2 AI laser pulse sequence for Ramsey-Bordé as opposed to the π/2-π-π/2 sequence for Mach-Zehnder (shown in Figure 2), QTEST's AIs can measure the quantity $h/m$ for $^{87}$Rb and $^{85}$Rb, where $m$ is the atom's mass and $h$ the Planck constant [18,19]. Combined with the Rydberg constant and the isotopes' mass ratio with the electron ($m/m_e$), it can determine the fine structure constant $\alpha$. By defining the Plank constant $h$, it can lead to a precise definition of SI mass unit based on atomic systems.

The sensitivity to $h/m$ of Ramsey-Bordé AIs comes from the asymmetric velocity patterns of the two interferometer paths. The asymmetry leads to higher mean kinetic energy of atoms traversing through one interferometer arm than through the other arm, thus the AI has first order sensitivity to the photon recoil energy, $hk^2/2m$, where $k$ is the wavenumber of the AI laser.

*2.2.1 Measuring h/m and the fine structure constant*

The fine structure constant is ubiquitous in physics. For example, the energy change of an electron bound within a hydrogen atom relative to a free electron is $-m_e c^2 \alpha^2/2$, the fine structure is smaller by another factor of $\alpha^2$, and the Lamb shift by yet another $\alpha$. The fine structure constant thereby sets the structure and hierarchy of matter. Precise knowledge of the fine structure constant will impact many fields of science. Today's best value – with a precision of 0.25 parts per billion – is derived from a measurement of the gyromagnetic anomaly $g_e$-2 of the electron [73] and its theory in terms of $\alpha$, summing up over tens of thousands of Feynman diagrams in the theory of quantum electrodynamics [24]. Unfortunately, $\alpha$ is not yet known from independent experiments to the same precision. Thus, what could be the most precise test of QED is hampered by our lack of knowledge of $\alpha$. QTEST will help measuring $\alpha$ to $10^{-11}$ precision without resorting to high-order quantum electrodynamics.

Comparison of such a precise measurement of $\alpha$ to other measurements would constitute one of the most precise tests of the overall consistency of the accepted laws of physics and experimental methods. In particular, it will allow the most precise test of the theory of quantum electrodynamics and establish a limit on a possible inner structure of the electron [74]. If $g_e$ doesn't deviate from the expected value by more than $\delta g_e$, the energy scale of such a substructure $m^* > m_e/(\delta g_e/2)^{1/2}$ (in the "chirally invariant" model – other models lead to a linear scaling and thus a larger scale). Current data yields $m^* > 0.7$ GeV, limited by measurements of $\alpha$. Data from the large electron-proton collider sets a limit of 10 TeV, but QTEST could reach 5-50 TeV, assuming equal progress in the measurement and theory of $g_e$. Thus, paradoxically, some of the lowest-energy experiments in physics yield some of the highest-energy bounds on elementary-particle substructure.

The measured value [75] of the muon's $g_\mu$-2 is more than 3 standard deviations (or 0.6 ppm) below standard model calculations. If the discrepancy is real, there should be a corresponding $1.5 \times 10^{-11}$ discrepancy for the electron (lower by the muon/electron mass ratio squared). To measure that, we must have correspondingly good knowledge of $\alpha$. This is within QTEST's reach.

*2.2.2 Absolute atomic masses*

The kilogram is the last unit that is defined by an artifact. This has obvious technical disadvantages, such as susceptibility to contamination or damage of the prototype. It also runs counter to the ideal of units based on nature's laws that can be reproduced equally well by any nation. In 2011, the General Conference on Weights and Measures expressed its intent to revise the definition by assigning an exact value to the Planck constant $h$ [76]. The kilogram would then be referenced to the second through the defined values of the Planck constant and the speed of light.

Based on this definition, QTEST will establish $^{87}$Rb and $^{85}$Rb as calibrated microscopic mass standards that can be used anywhere in the world, based on measuring $h/m$ ($^{87}$Rb) and $h/m$ ($^{85}$Rb) [25]. These QTEST-standards will have a precision of $10^{-11}$. They will enable absolute microscopic mass measurements with unprecedented precision, greater than 1000 times more accurate than that in the present SI. [77] Other microscopic masses can be related to the QTEST-standard atoms by mass spectroscopy. The link to macroscopic masses can be made by Avogadro spheres: silicon crystals of accurately measured atom number [78]. QTEST's mass standards work in the same framework as the Watt balance [79,80]. They are based on inertial mass and do not rely on Earth's gravity (which varies with location and time due to tides, earthquakes and magmatic currents), do not require mechanically moving parts or standard resistors that are prone to drift, are based on fundamental laws of quantum mechanics, rather than macroscopic quantum effects (for which first-principles theory doesn't exist) and realize high precision in the microscopic world, where it is most needed – measuring macroscopic masses is limited by, *e.g.,* contamination, chemical activity, or the buoyancy of air, but microscopic masses can be compared to $10^{-11}$ precision or better [81].

## 3 QTEST Concept Design Summary

### 3.1 Differential measurements for EP test

AIs are highly sensitive to accelerations (*a*) with accumulated phase response in Mach-Zehnder configuration given by $\phi=(\bm{k}_{\text{eff}}\cdot\bm{a})T^2$, where $\bm{k}_{\text{eff}}$ is the effective wavevector and *T* is the free-evolution-time between interferometer pulses, whose gravity-induced evolution of their quantum state is measured by reading out the interference pattern. The phase measurement sensitivity is therefore determined by the signal-to-noise ratio (SNR) in detecting $\phi$ and by *T*. A fundamental limit for the SNR is the atomic shot noise, or quantum projection noise, which is scaled by the finite number of atoms participating in detection as $N^{-1/2}$. Reduction of the visibility of the interference fringe affecting SNR is attributed to inhomogeneity and imperfections in the atomic state preparation and detection. These factors are summarily parameterized by the contrast, $C = (R_{\max} - R_{\min})/(R_{\max} + R_{\min})$ with $R_{\max}$ ($R_{\min}$) being the maximum (minimum) normalized population in the excited state after the interferometer sequence. An analysis of the expected fringe contrast in the QTEST apparatus and imaging-based detection schemes for conserving signal contrast at high sensitivity are discussed in Section 4.6.

For the optimal system performance, strict control over the atomic cooling and initial state preparation must be maintained throughout the mission lifetime, limiting the practical cycle time to $T_C$ = 70 s, given by $2T$ = 20 s interferometry interrogation times and < 50 s for atomic cloud sample preparation, cooling, state manipulation, detection, and positioning of the atoms and the rotating platform. Assuming that each interferometer has a beam near-resonant to the $D_2$ transition in rubidium, with $k_{\text{eff}} = 2k$ where $k = 2\pi/(780\text{nm})$, contrasts in excess of 50%, and $N > 10^6$ detected atoms, the per-shot acceleration sensitivity for each rubidium AI is:

$$\sigma = \sqrt{\frac{1}{N}}\frac{1}{C*k_{\text{eff}}*T^2} = 1.3 \times 10^{-13}\ g. \tag{3}$$

The QTEST apparatus relies on differential measurements of two, simultaneously interrogated, atom interferometers ($^{85}$Rb and $^{87}$Rb) to search for relevant violations of the UFF. For a total measurement time of 12 months, allowing for a multiple of 3-month measurement sets for systematic reduction, the integrated acceleration sensitivity for each differential QTEST UFF measurement set is:

$$\sigma_\eta = \sqrt{\frac{T_C(\sigma_{85}^2+\sigma_{87}^2)}{\tau}} = 5.3 \times 10^{-16}g, \tag{4}$$

where $T_C$ is the measurement cycle time and $\tau$ the total time of the measurement set. Comparing to the state of the art, summarized in Table 2, it is anticipated that QTEST will provide two or more orders of magnitude improvement in constraining the Eötvös parameter over any previous test of the WEP.

| Experiment | Species/Comments | $T$ (s) | Accuracy [projected] |
|---|---|---|---|
| Torsion balance [82–84] | Ti, Be | | $10^{-13}$ |
| LLR [85] | Lunar Ranging | | $10^{-13}$ |
| **AI/FG-5** [86] | Cs, Glass | | **$7 \times 10^{-9}$** |
| **Wuhan** [17] | $^{87}$Rb/$^{85}$Rb | 0.071 | **$3 \times 10^{-8}$** |
| **Hannover** [16] | $^{87}$Rb/$^{39}$K | 0.02 | **$5 \times 10^{-7}$** |
| **Hannover VLBAI** [87] | $^{87}$Rb/$^{170}$Yb | 1.3 | **[$7 \times 10^{-13}$]** |
| **Stanford** [88,89] | $^{87}$Rb/$^{85}$Rb | 1.34 | **[$10^{-15}$]** |
| **SAI** [11] | $^{87}$Rb/$^{85}$Rb or Rb/K | 1 | **[$10^{-12}$]** |
| **ICE** [90] | $^{87}$Rb/$^{39}$K | 0.01 | **[$10^{-11}$]** |
| **Quantus** [91] | $^{87}$Rb/$^{41}$K | 1 | **[$6 \times 10^{-11}$]** |
| **STE-Quest** [92] | $^{87}$Rb/$^{41}$K | 5 | **[$10^{-15}$]** |
| **QWEP** | $^{87}$Rb/$^{85}$Rb | 1 | **[$10^{-13}$]** |
| Microscope [93] | | | [$10^{-15}$] |
| STEP [94] | Be/Nb | | [$10^{-18}$] |
| GG [95] | | | [$10^{-17}$] |
| SR-POEM [96] | Sounding Rocket | | [$2 \times 10^{-17}$] |
| **QTEST** | $^{87}$Rb/$^{85}$Rb | 10 | **[$5 \times 10^{-16}$]** |

**Table 2:** Leading and proposed tests of the WEP. Experiments with quantum test-masses are distinguished in bold. In comparison with the numerous proposals (projected accuracy in brackets), the QTEST mission will utilize recent developments in space-based quantum sensors for unprecedented WEP tests with quantum test-masses.

### 3.2 Advantages of atom interferometer WEP measurements in space

ISS offers a unique space platform for ultra-cold atom experiments in general and atom interferometry in particular. It has been well-recognized that microgravity in space allows long interrogation times for unprecedented AI sensitivity [97]. Equally important is the device size requirements that are limited only by the atom's drift velocity and the splitting of the atomic trajectories, on the order of several tens of centimeters for $T$=10 s, allowing for better local control of the atoms' environment. This is in contrast to the few 100 meters of free fall distance required for terrestrial AIs with the same $T$. Further, the absence of the large gravity bias is ideally suited for advanced cooling schemes that are necessary to achieve ultra-high precision at long $T$ [8,13,98,99].

QTEST will critically rely on the ability to reverse the gravity relative to the instrument measurement direction on ISS. As will be discussed in the following sections, there exist many systematics that are nearly impossible to otherwise eliminate. Many of the largest systematic effects stem from imperfect overlap of the two atomic species and the existence of the gravity gradient, that always accompanies the presence of the gravity field. Reversing the gravity vector relative to the AI measurement direction will change the resulting systematic effect signal signs with respect to that of the WEP signal. By modulating the gravity direction, therefore, QTEST will be able to reduce the systematics by averaging rather than being overburdened by addressing each systematic effect with brute force.

### 3.3 Unique approaches in QTEST

QTEST uses dual-species atom interferometers and their differential measurements for testing WEP in space, similar to previously proposed mission concepts (see Table 2), namely the early NASA QuITE, and ESA QWEP and STE-QUEST [92,100,101]. However, QTEST employs unique approaches to reduce or eliminate systematics and to achieve the requisite measurement precision. It will be the first to take advantage of the space environment and actively modulate the gravity signal by a rotating platform mount on ISS. It also integrates well-established evaporative cooling techniques with delta-kick cooling to achieve ultra-low kinetic energies in highly-uniform quadrupole-Ioffe configuration (QUIC) magnetic traps [102,103]. Finally, it utilizes the recent advances in Bragg-diffraction atom optics and imaging detection for enhanced signal while minimizing systematic errors.

Mounting the AIs on a rotating platform allows for reversal of gravity relative to the orientation of the apparatus, thereby modulating the differential phase for two rubidium AIs according to $\Delta\phi = \Delta\phi_g \cos(\omega t) + \phi_{syst}$, where $\Delta\phi_g$ is the signal from differential gravitational accelerations of the two gases and $\phi_{syst}$ are from external systematics that are not correlated with the rotating platform modulation frequency ($\omega$). Averaging over a time $\tau$ by means of a product detector demodulates the gravity signal to DC while all remaining systematics are averaged down, in a simple analogy to a lock-in detector:

$$\frac{2}{\tau}\int_0^\tau \Delta\phi \cos(\omega t)dt \leq \left|\Delta\phi_g + \frac{2}{\omega\tau}\phi_{syst}\right|. \tag{5}$$

For example, if gravity is reversed every 10$^{th}$ measurement, for a 3-month integration time, the reduction factor for uncorrelated systematics will be better than $10^4$. Gravity reversal is uniquely applicable in microgravity environments and offers one of the most effective techniques for reducing measurement systematics in QTEST. The rotating platform operation allows for the modulation at any frequency, e.g. asynchronous with ISS rotation and orbit and can thus be utilized for making QTEST measurements independent of ISS platform rotation and for suppression of the Sagnac and Coriolis forces. As discussed in Section 5, even the conservative suppression factor assumed above well-satisfies the most stringent requirements for testing the WEP to below a part in $10^{16}$. Further, ultralight dark matter candidates such as B-L gauge bosons might give rise to time-dependent EP violations, QTEST is the first experiment purpose-designed to probe for such effects.

QTEST will use Bragg-diffraction atom-optics. The main motivation for the choice is the possibility of using a single Bragg laser system for $^{85}$Rb and $^{87}$Rb with a single, far-detuned laser addressing both rubidium isotopes. It is relatively easy to find a "magic" wavelength that addresses both species with the same Rabi frequency, and makes perfect wavenumber matching possible. This eliminates the influence of vibrations and laser phase noise completely in principle. Bragg diffraction does not change the atoms' internal quantum state, which will help to reduce systematic effects, *e.g.*, magnetic fields and AC Stark shifts. The single Bragg beam is thus an enabling approach for ISS platform, as ISS is not a vibration-isolated platform. It should be pointed out that a parasitic phase shift was observed with Bragg diffraction in simultaneous conjugate interferometers [25,104] that used to be of concern for the precision measurements. It is now well understood as it has been explained quantitatively by solving the Schrödinger equation for the atomic matter wave in a laser field. The simulations have been confirmed experimentally [105]. For Mach-Zehnder interferometers, as we will use them in the WEP test, symmetry allows for measurements in which the phase shift can in principle be eliminated completely.

QTEST will use QUIC-trap-based ultracold atom sources to yield the maximum achievable populations of both Rb species with high spatial symmetry. Two separate dual atom-sources are designed for simultaneous operation of the two dual-species AIs with opposite sign of $k_{eff}$ for simultaneous k-reversal measurements [106], which not only reduces systematics scaling as $k_{eff}^0$ and $k_{eff}^2$, but also provides redundancy in space in that each source alone is sufficient for a test of the WEP with unprecedented, although reduced, precision and the full set of the planned photon-recoil-based measurements. Delta-kick expansion will be used to achieve low-density ($n_0$), low-temperature samples for long free-expansion times with minimal collisions in a compact, well-controlled environment. This cooling technique relies on the position/momentum correlation for atoms after ballistically expanding, with slower atoms occupying the center of the gas and faster atoms traveling to the edges after sufficient expansion time. Application of a position-dependent impulse, with sufficient strength and harmonicity of the spatial profile, then drastically reduces the expansion velocity of all atoms in the gas.

Even with delta-kick produced atomic clouds at nanoKelvin effective temperatures, the practical interrogation time is limited due to the Coriolis force and imprecision in the cloud overlaps, mostly due to the reduction of the AI fringe contrast. Two new developments recently reported and demonstrated will help address this limitation. As detailed in Section 4.6, imaging detection and an "open interferometer" scheme, using controlled asymmetry in the interferometry timing, will not only recover interferometer contrast in the presence of gravity gradients and rotations, but also relax the WEP-test constraints for dual-species overlap by orders of magnitude.

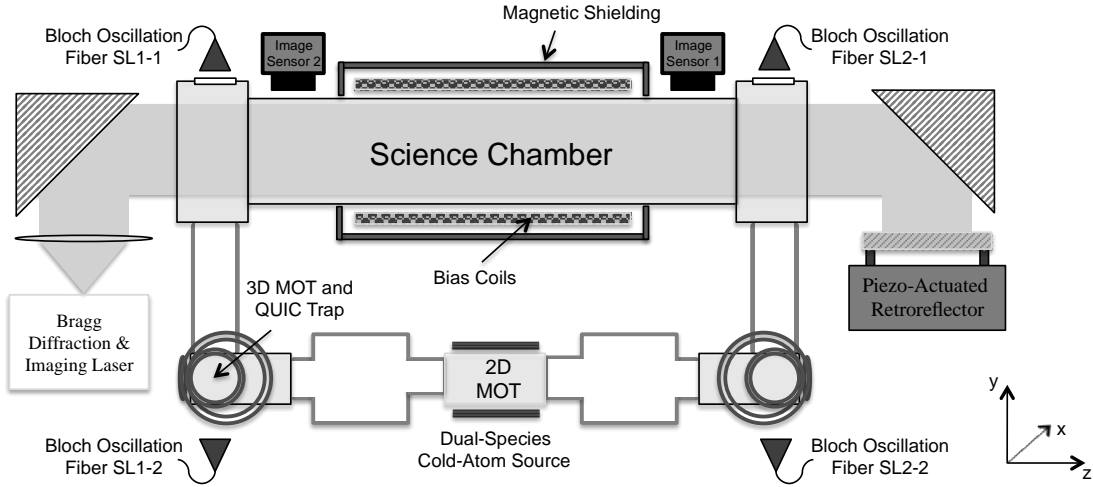

**Figure 1:** QTEST AIPP, including two dual-species atom sources, fiber-delivered optical lattice transport via Bloch oscillations, the Bragg-based AI system designed to simultaneously address four gas clouds (initially overlapped $^{87}$Rb and $^{85}$Rb at each side of the magnetically-shielded science chamber), and the dual-species imaging systems for AI phase readout. SL indicates the starting points of the ultracold atom clouds.

## 4 Experiment

The QTEST experiment payload will mainly consist of the atom interferometer physics package (AIPP), a rotating platform, a laser and optics subsystem assembly, and control and electronics.

The schematic concept of the AIPP is illustrated in Figure 1. Here, a dual-species ($^{85}$Rb, $^{87}$Rb) 2-D magneto optical trap (MOT) feeds two identical atom-source regions at opposing sides of the vacuum chamber. At each, the cold Rb beams are collected and simultaneously laser-cooled in 3-D MOTs and subsequently evaporative ($^{87}$Rb)/sympathetically ($^{85}$Rb) cooled in a QUIC magnetic trap [102,103]. A final delta-kick cooling stage [7,13] and state-transfer to the magnetically insensitive ($m_F = 0$) state will reduce and maintain the kinetic energy of each atomic species, at each trap, to be consistent with an effective temperature of less than 1 nK with $10^6$ atoms.

After the complete cooling process, the ultra-cold atom clouds are transported to the science chamber (SC), where AI measurements take place in a well-controlled and isolated environment. An optical Bloch oscillation (OBO) technique will be used to transport the ultra-cold atoms to SC. OBO is used not only for its ability to transport atom ensembles without heating [107], but also with large momentum for reduced transport time. OBO of cold atoms in a moving optical lattice has been demonstrated for transferring >1000 $\hbar k$ photon momenta in 10 ms for precision metrology [18,107]. The transfer efficiency is only limited by spontaneous emission loss. Overall, we will show that we can transport the dual species atom ensembles from the ultracold atom sources to the science measurement region without sacrificing atom numbers, temperature, and overlap between the two atom clouds.

Interferometry for the two species will be performed with a sequence of Bragg laser pulses. A typical π/2-π-π/2 sequence makes a Mach-Zehnder interferometer configuration highly sensitive to accelerations. The Bragg pulses will be shaped in a Gaussian time-profile to improve the atom optics efficiency. With the planned interferometer interrogation time of $T$=10 s, a mere $4 \times 10^{-10}$ g of effective acceleration will result a full fringe shift. Therefore, spatially inhomogeneous forces, including gravity gradients and rotations, can be sufficiently large to reduce the total contrast to zero. To mitigate this issue and still have the ability to take advantage of the long interrogation times accessible in space, QTEST will use open interferometry [9,13,108] to compensate the inhomogeneous effects and imaging detection [7,9,13] to recover the contrast loss due to the inhomogeneity. These techniques have been proven for recovering spatially dependent phase shifts in long-time AIs that would otherwise have been completely washed out with state-detection over the entire clouds.

In the following sections, we will describe in detail the major design considerations of the QTEST concept for enabling the stated science objectives and for mitigating adverse effects unavoidable on ISS.

### 4.1 Atom interferometer

Figure 2 shows a space-time diagram illustrating the two counter-propagating dual-species atom interferometer trajectories for the WEP measurements. Initially, two samples of both rubidium isotopes, overlapped at the starting locations SL1 and SL2, respectively, are launched with a π Bragg-pulse at a drift velocity of $\pm\hbar k_{\text{eff}}/m$. A common laser is used at the "magic" wavelength so that the two-photon Rabi frequencies for both isotopes are equal. Since the recoil velocity is 87/85 times higher for the lighter isotope, the species separate thereafter. At about 13 seconds drift time, when the atoms are fully in the shielded drift tube region of the SC, the π/2–π–π/2 sequence of the Mach-Zehnder interferometer starts. The two outputs of the $^{87}$Rb interferometers reach each detection region at 38.6 and 44.2 seconds, respectively; the lighter species arrives about 1 second earlier. The single laser pulse for establishing drift velocities of the two samples in opposite directions utilizes the degeneracy of the two initially stationary atom samples. In the process, however, half of the atoms are lost.

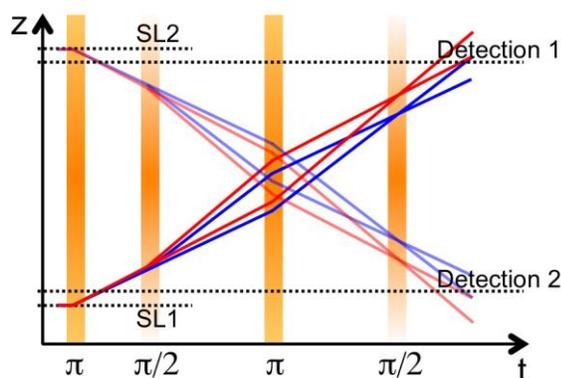

**Figure 2:** Space-time diagram of the atom interferometers. The difference between the paths taken by $^{87}$Rb (blue) and $^{85}$Rb (red) is exaggerated.

The atom interferometer pulse separation time of $T$=10 s has been chosen based on a trade-study to optimize the signal and precision targeted while minimizing instrument design complexity, power requirements, and size of the physics package. The length of the overall SC of approximately 70 cm is sufficiently compact for the rotating platform while still achieving the required sensitivity (see Section 5). Large-momentum transfer (LMT) schemes [10,109] are often considered as a way to improve the sensitivity of the atom interferometer. It turns out that, counter-intuitively, for a given physical size of the atom interferometer region, simple two-photon momentum transfer with longer interrogation time offers higher sensitivity than LMT. Therefore, LMT is not planned in QTEST.

An alternative geometry, the symmetric diamond configuration, has also been considered. It offers the opportunity to suppress several systematic effects [110] related to the gravity gradient. However, diamond-configuration AIs are prone to beam splitting loss and require clearing pulses during interferometry. They are also not maturely proven and require more complicated phase readout with three coupled AI output ports.

### 4.2 Bragg laser and retroreflecting mirror

A single Bragg laser addressing both rubidium isotopes guarantees that this differential measurement has the highest common mode rejection to vibrations on ISS. For this to work properly, the Bragg laser will have to provide the same beam splitting and combining functions for both species. Fortunately, a magic wavelength can be found where the two-photon Rabi frequencies for $^{85}$Rb and $^{87}$Rb are identical. The two photon transition $|0\rangle \leftrightarrow |2\hbar k\rangle$ has a resonance frequency of $4\Omega_r = 2\hbar k^2/m$. Due to the small transition frequency difference between $^{85}$Rb and $^{87}$Rb of $\simeq 2\pi \times 355$ Hz, compared to the bandwidth of a typical Bragg diffraction of pulse duration $\sim 100$ μs, simultaneous transitions of both species using a single frequency pair is feasible. For the proposed AI configuration, the two-photon transitions are between $|2\hbar k\rangle \leftrightarrow |4\hbar k\rangle$, which increase the frequency difference between species by a factor of 3 to $2\pi \times 1.065$ kHz. Figure 3 shows the relative difference in the two-photon Rabi frequencies ($\Omega$) versus the Bragg beam wavelength for the $m_F = 0$ (clock) states, where the zero crossings indicate "magic" wavelengths for the pair of internal states chosen for QTEST. The pairs of interest are $^{85}$Rb F = 3, $^{87}$Rb F = 2, and $^{85}$Rb F = 2, $^{87}$Rb F = 1, for which the Zeeman shifts partially cancel.

The choice of the single Bragg laser approach for the dual species interferometers also eliminates the differential measurement sensitivity to atom interferometer laser phase noises, much the same way as for vibrations. There is no need for precise phase locking of separate lasers and precise wavelength ratio control as is necessary when two very different wavelength lasers are used, *e.g.*, in the Rb and K measurements [111]. It is important to note that use of two frequency-components reflected off a common retro-reflecting mirror for Bragg diffraction introduces the wavelength $k$ vector reversal symmetry, making |i⟩ ↔ |i+2ℏk⟩ transitions indistinguishable from |-i⟩ ↔ |-(i+2ℏk)⟩. Hence, the single Bragg laser with two retro-reflected frequency components will simultaneously address all four clouds in the interferometer with identical $k_{eff}$ and Ω, accomplishing both dual species interferometry and $k$-vector-reversal measurements all with a single Bragg pulse sequence.

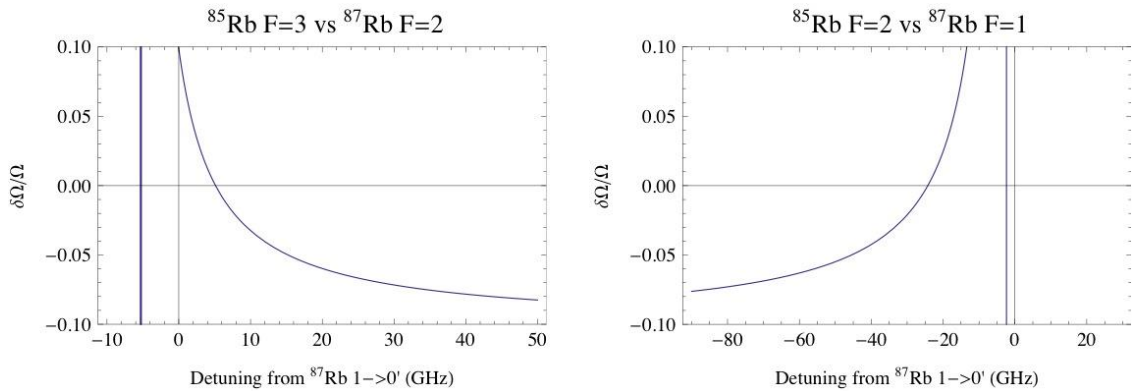

**Figure 3:** Relative two-photon Rabi frequencies of $^{85}$Rb and $^{87}$Rb versus optical frequency. Fractional difference $\delta\Omega/\Omega$ of the Rabi frequency for two-photon Bragg diffraction of the $m_F$=0 state is calculated based on the Clebsch-Gordan coefficients of the D$_2$ transition hyperfine structure.

Requirements for the Bragg laser are relatively easily met. In one analysis, the laser can be a diode laser amplified with a 2-W tapered amplifier. An acousto-optical modulator after the amplifier shapes the Gaussian Bragg pulses with a duration of several 100μs, *e.g.* a temporal profile of exp(-$t^2$/2$\tau^2$), where $\tau \sim 66$ μs [109]. The Bragg laser will be collimated at a $1/e^2$ radius (waist) of 24 mm. This radius is chosen to be at-least twice as large as the Gaussian width of the atomic clouds after thermal expansion of 20 s interferometer total interrogation time. Simulations of the expected signal in this system predict a minimal contrast loss (~10%) due to the intensity-inhomogeneity of the Bragg beam across the atomic cloud.

**4.3    Atom sources**

For simultaneously accommodating the operation of k-reversal interferometers, the QTEST instrument is designed to have two dual species sources, each species contains a minimum of $10^6$ atoms at the final preparation stage with a temperature less than 1 nK. However, cooling atoms to Bose-Einstein condensation (BEC) is not necessary. This avoids the considerable difficulty of condensing $^{85}$Rb, on the one hand, which suffers from large 3-body losses at high densities and requires Feshbach tuning of the s-wave scattering length in an optical dipole trap for condensation [112]. On the other hand, mean field shifts are typically very large while atom numbers are low in BECs, and BEC preparation times are long, often making degenerate gases unfavorable for precision metrology. For QTEST, ultracold atoms at low densities at the time of interferometry are desired.

*4.3.1    Atom collection*

Atoms are initially prepared in high vacuum from a rubidium dispenser in a dual-species 2D-MOT that is directed through differential pumping apertures to the ultra-high vacuum 3D MOT regions [103]. For $^{87}$Rb, $N > 10^9$ atoms cooled to 25 μK can be expected after loading the 3D MOT for 1 second, and a few milliseconds of polarization gradient cooling [113]. For $^{85}$Rb, we will only collect and cool a few $10^6$ atoms in optical molasses for the sympathetic cooling stage.

In order to meet the interferometer SNR requirements, QTEST will use a QUIC trap source instead of a chip-based ultra cold atom source. While the latter has the distinct advantages of lower power requirements and high rethermalization rates for rapid evaporative cooling, the technology is less mature and the trap profiles are less favorable for delta-kick cooling. In fact, only one chip-source has recently demonstrated loading of sufficient numbers of $^{87}$Rb atoms from a MOT [114]. The chip trap has a large asymmetry to achieve the necessary large volume, making the subsequent delta-kick cooling difficult. In comparison, the technology of sympathetic cooling of $^{85}$Rb with $^{87}$Rb in QUIC traps is well-over a decade old [102] and large trap depths with controllable trap symmetry are intrinsic to the systems. In the following section, we discuss a compact design to realize the QUIC-trap within the power and thermal constraints of ISS.

*4.3.2    Evaporative cooling in a QUIC trap*

Laser cooling atoms in an optical molasses is practically limited to ~μK temperatures and densities on the order of $10^{10}$ - $10^{11}$/cm$^3$ limited by the photon recoil and optical depths of the clouds [115]. To reach 1 nK temperature at sufficiently high density, two additional cooling stages are needed: evaporative cooling in a magnetic trap followed by a final delta kick cooling. After the dual-species laser cooling stage, the expected phase-space density is on the order of ~

$10^{-6}$. In comparison, the final clouds are at a phase space density of $> 10^{-4}$, assuming $\sigma_r = 6$ mm and temperatures $< 1$ nK. Evaporative cooling will bridge the gap. For the evaporative cooling stage, the atoms are first optically pumped (with an applied ~10 mG bias magnetic field) into the weak-field seeking states $|F = 2, m_F = 2\rangle_{87Rb}$ and $|F = 3, m_F = 3\rangle_{85Rb}$ and then loaded via magnetic transport and adiabatic compression into a QUIC trap [116], consisting of two identical quadrupole coils and one additional Ioffe coil.

After the atoms are loaded into the QUIC trap, sympathetic cooling of $^{85}$Rb with $^{87}$Rb is applied to increase the phase-space density before delta kick cooling. Sympathetic cooling of the two gases in thermal equilibrium is characterized by the $\alpha$ parameter as [117]:

$$\frac{T_i}{T_f} = \left(\frac{N_{87Rb}}{N_{85Rb}} + 1\right)^{\alpha}, \tag{6}$$

where $\alpha > 0.77$ for our dual-species rubidium gases (extracted from data in Ref. [103]), $T_i$ ($T_f$) are the initial (final) temperatures of the dual-species gases in thermal equilibrium, and $N_{87Rb}$ ($N_{85Rb}$) are the majority (minority) atomic species. Sympathetic cooling is assumed to progress until the populations are equal.

With the QUIC traps in a space platform, electric power dissipation is of concern. The design of the source vacuum chamber and the magnetic coils must take this factor into consideration. The spatial profile of the magnetic field inside the source cell, from the QUIC trap, was numerically simulated to optimize the atom confinement, power requirement, and harmonicity of the trap. With reasonable coil size choices, we find that two identical quadrupole coils carrying 25 A, requiring 270 W total power, and one Ioffe coil carrying 25 A, with 120 W total power, can create a magnetic trap with $U_0 = k_B \times 10$mK depth, where $k_B$ is Boltzmann's constant, and trap frequencies $\omega = 2\pi \times$ (28 Hz, 24 Hz, 26 Hz). Table 3 summarizes the key parameters achievable for the atomic gasses. The second and third columns of Table 3 list the estimated efficiencies of the envisioned MOT/QUIC trap loading and evaporative/sympathetic cooling stages, based on demonstrations in similar lab-based systems [103]. The relatively large magnetic-field minimum in the trap ($B_0$~230 G) maintains the atoms in the weak-field-seeking "stretched" states. Thus, they have the same linear Zeeman shift of 1.4 MHz/G so that the trap potentials for both species are the same.

*4.3.3 Delta-kick cooling*

After evaporative/sympathetic cooling, the atoms are prepared as near point sources for delta-kick cooling to ultra-low temperatures below 1 nK. Cooling is defined here in the weakest sense; the atoms are neither in thermal equilibrium nor benefit from enhanced degeneracy at the end of the delta kick cooling stage. In fact, it is a process that simply reduces the mean kinetic energies by sacrificing density at constant phase space density [98,99,118]. This cooling scheme is ideally suited for the high symmetry of microgravity to extend the free fall times without atom clouds expanding to unmanageable sizes.

| | **QUIC Trap Loading** | **End of Evaporation** | **After Delta Kick Cooling** |
|---|---|---|---|
| $N_{87Rb}$, $N_{85Rb}$ | $3 \times 10^8$, $2 \times 10^6$ | $> 10^6$ each | $> 10^6$ each |
| Temperature | 100μK | 2.5μK | 1nK |
| Peak Density ($^{87}$Rb) | $8 \times 10^{10}$/cm$^3$ | $7 \times 10^9$/cm$^3$ | $4 \times 10^5$/cm$^3$ |
| Phase Space Density ($^{87}$Rb) | $10^{-7}$ | $10^{-4}$ | $10^{-4}$ |

**Table 3:** Modeled performance for cooling $^{87}$Rb and $^{85}$Rb clouds from 3D-MOTs to free-space, dual-species gases after delta kick cooling.

|  | **X-Axis** | **Y-Axis** | **Z-Axis** |
|---|---|---|---|
| $B^{(3)}/B^{(2)}$ | $10^{-16}$ | $2.2*10^{-16}$ | 1.71 |
| $B^{(4)}/B^{(2)}$ | .090 | 2.1 | .58 |

**Table 4:** Dominant anharmonic terms for the simulated QUIC trap B-field profile along the Quad-coil axis (x), along the Ioffe-coil axis (z), and transverse to both coil axes (y). All terms satisfy the requirements for delta-kick cooling to 1nK by a large margin.

At the end of the evaporation, the atoms will be transferred to the $m_F = 0$ state by Raman or microwave pulses. This effectively turns off the trap and leaves the atoms free to expand without being affected by the magnetic trap field. The magnetic fields can even be ramped off/on during this time. After the requisite ballistic expansion, ~400 ms at 2.5 μK initial cloud temperature, the atoms are then driven back to the stretched states. The spatial dependence of the impulse experienced by the atoms while they are in the stretched states is designed to match the cloud energy distribution and thus minimize their energy. The ballistic expansion time is optimized separately for each of the two species, which is possible because their hyperfine-state transitions are energetically well resolved. At the end of the delta kick cooling scheme, the atomic clouds each have a 1/*e* radius of ~5 mm and a kinetic energy of $< k_B \times 1$ nK. The projected properties of the atomic clouds after all cooling stages, just before transport into the SC for interferometry measurements, are given in the final column of Table 3.

Details of the scaling and tolerances for delta kick cooling with trap-field, expansion time, and pulse times have been well summarized in the Science Envelope Requirements Document for the Cold Atom Laboratory [119]. The primary concern for achieving ultra-low temperatures in this cooling scheme derives from anharmonicities of the magnetic trap, which will limit the ultimate temperature and induce center-of-mass motion of the clouds. The simulated magnetic trap is nearly harmonic at the center, so that only the first few terms in the Taylor-expansion of the field profile are significant:

$$B(x) \cong B^{(0)} + \frac{1}{2}B^{(2)}x^2 + \frac{1}{6}B^{(3)}x^3 + \frac{1}{12}B^{(4)}x^4. \qquad (7)$$

Constraining the influence of the forces from higher-order harmonic terms to be smaller than the final momentum spread of the cloud yields:

$$\frac{B^{(n)}}{B^{(2)}} < \left(\frac{\sigma_i}{\sigma_f}\right)\frac{(n-1)!}{\sigma_f^{(n-2)}}, \qquad (8)$$

where $\sigma_i/\sigma_f$ is the ratio of the initial ($\sigma_i$) to final ($\sigma_f$) atomic cloud sizes for the delta kick cooling ballistic expansion stage and n is the Taylor-expansion-order. QTEST requires $B^{(3)}/B^{(2)} < 13.5$ and $B^{(4)}/B^{(2)} < 20250$. Fitting the simulated QUIC trap field profiles to the 4th order polynomial, given in Table 4, demonstrates that the anharmonicity of the simulated QUIC trap is negligible for all directions.

### 4.4 Atom transport

Separation of the atomic sources and the SC allows QTEST to optimize the control and isolation of the science measurement environment while maximizing the source throughputs. The SC will have enhanced optical access for the Bragg beams and high suppression from potential sources of light, vacuum, and magnetic-field perturbations. The added cost is the atom transport process. A viable atom transport method should meet the following requirements: short transport time (< 1s) to minimize excess thermal expansion of clouds, have high transport efficiency (~100%) to preserve atom flux, and preserve the positioning and overlap repeatability of the two species without inducing additional systematics. We propose to use Bloch oscillations (for fast transport) in a moving optical lattice to displace stationary clouds from the delta kick cooling region to the center of the SC. This is accomplished by counter-propagating laser beams, with control of their amplitudes and frequency differences.

Moving optical lattices are formed by the interference of the counter-propagating laser beams from Bloch oscillation fiber outputs SL1-1 and SL1-2 and fiber, SL2-1 and SL2-2 respectively (shown in Figure 1). These beams are adiabatically turned on and off in less than 1 ms [18]. The moving lattice constantly accelerates and then decelerates to rest by changing the frequency difference of the two lasers [18,107]. The clouds are thus transported away from the delta kick cooling region, reaching the axis of the SC and ready for the atom interferometer sequence. Following Ref. [120], we find that, with realistic laser and TA systems, the cloud can be transported with separate beams over 10 cm in 0.5 s with 90% efficiency.

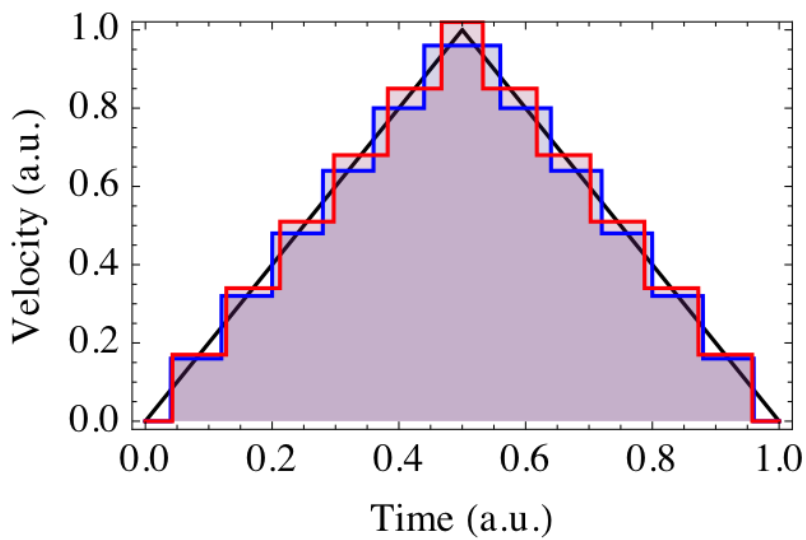

Due to the difference in mass, $^{85}$Rb and $^{87}$Rb will have different velocities after receiving identical photon momenta. A previously proposed scheme uses the same lattice but different photon-momentum transfer for $^{85}$Rb and $^{87}$Rb to closely match the final velocity [88,89] for coherent launch. However, for pure transportation without net momentum transfer from stationary to stationary states, it is possible to maintain the spatial overlap of the species. In a lattice moving at constant velocity, trapped atoms follow the classical trajectory defined by the lattice wells together with the quantum nature of discrete photon momentum transfer, which displaces the two species after transport as illustrated in Figure 4. By designing the right sequence for the instantaneous velocity of the lattice, one species of choice can stay at higher velocity than the other for a defined duration. The final overlap of the two species is therefore controllable with the stability determined by electronics. This picture applies both to deep lattices and shallow lattices, where the actual velocity profile can be tuned up experimentally in operation. Therefore, this scheme may offer a handle for fine-tuning the spatial overlap of the two species along one axis.

**Figure 4:** Depiction of velocity evolution of two species undergoing Bloch oscillations in a common moving optical lattice. Black line: lattice velocity. Red (blue) curve: ensemble-averaged velocity of $^{85}$Rb ($^{87}$Rb). The areas of the shaded regions under the curves demonstrate equal displacements of the two atom clouds. In the weak lattice limit, the velocity of the atoms is discrete.

### 4.5 Two-species overlap

High precision AIs are critically sensitive to environmental perturbations and externally applied forces. The extensive common-mode rejection built into the QTEST concept mitigates much of this susceptibility. Most notably, the use of identical $k_{eff}$ with Bragg interferometry of the two isotopes of rubidium, and simultaneously operated dual-species AIs with equal but opposite momentum states traversing overlapping classical paths will result in the leading-order phase shifts (*e.g.*, from translations, rotations, and gravity gradients) to be common between the two species and, in principle, perfectly cancel in analysis. In practice, very slight offsets in the center of mass (CM) position ($\delta r$) and velocity ($\delta v$) of the $^{87}$Rb and $^{85}$Rb test mass clouds break the commonality and symmetry and will lead to sizable shifts (see Section 5.1). It should be noted that the constraints of the CM overlap of the test-masses are general for all WEP type of tests due to the prevalence of the gravity gradient [121].

The CMs of two species actually cannot overlap over the entire trajectory except at one point in a typical MZ interferometer configuration, as shown in Figure 2. This is due to the inevitable mass difference between the two atomic species. In the presence of a gravity gradient, therefore, a differential interferometer measurement, without any compensation, will always include a phase error due to the gravity gradient even when the cloud CMs are initially perfectly overlapped. Removing this phase error requires precise knowledge of the gravity gradients, which is nearly impossible. Nevertheless, consistent initial overlap conditions in both position and velocity ensure that such phase error is constant from measurement to measurement. QTEST addresses this issue by using gravity reversal and k reversal measurements in which the displacement-dependent gravity gradient phase terms change sign relative to the possible WEP signal. Any gravity reversal induced displacement bias must be below $\delta r \sim 34$ nm and $\delta v \sim 0.4$ nm/s (See Section 5.1) so that the combined measurements can reach below $1 \times 10^{-16}$ g.

Shot-to-shot noise fluctuations between subsequent interferometer runs also add noise. In order not to degrade the SNR from the dominant gravity gradient dependent terms, maintaining high-stability in $\delta r$ and $\delta v$ is still critical. This requires the statistical CM fluctuations $\delta r_{Stat} < 34\mu$m and $\delta v_{Stat} < 0.4\mu$m/s to not exceed the atom shot noise (see Table 8). For our Gaussian clouds of 6mm width at 1nK, statistical CM fluctuations attributed to the finite atom numbers are similar for the two species, on the order of $\delta \sigma_r \sim 6\mu$m and $\delta \sigma_v \sim 0.3\mu$m/s, within the required stability.

## 4.6 Contrast loss and its mitigations

The standard method for extracting the AI phase shift relies on comparing the populations of atoms in each AI output port, yielding interference fringes. Imperfections in state preparation, atomic cloud inhomogeneity, and force gradients can reduce the achievable contrast, limiting the ultimate AI sensitivity. This limitation is particularly severe in space experiments, where one wants to take advantage of available long free fall times for ultra-high sensitivity AI measurements. For long atom interferometer interrogation times, gravity gradient and rotational phase-shifts across the length scale of the atomic clouds will induce sizable phase gradients and can easily make the contrast drop to zero [108,122]. Table 5 gives the magnitudes of the dominant phase shifts across the Gaussian width of the clouds in position ($\sigma_r$) and velocity ($\sigma_v$) for the ballistically expanded $^{85}$Rb and $^{87}$Rb gases in the proposed QTEST measurement environment. Considering the dominant terms, we would expect multiple phase-fringes on each interferometer output port, rendering the standard AI phase extraction protocol useless. The following sections discuss the proposed approaches in QTEST to mitigate contrast loss and to extract the WEP signal.

| Phase Term | $\Delta\phi$ ($^{87}$Rb) | $\Delta\phi$ ($^{85}$Rb) | Comments |
|---|---|---|---|
| $k_{eff} T_{zz} \sigma_r T^2$ | 25.1 | 25.1 | Gravity gradient (initial position) |
| $k_{eff} T_{zz} \sigma_v T^3$ | 13.0 | 13.1 | Gravity gradient (initial velocity) |
| $-2 k_{eff} \sigma_v \delta\Omega_y T^2$ | -1.14 | -1.15 | Coriolis |
| $-\dfrac{k_{eff}}{2 R_{eff}} (\sigma_r + \sigma_v T)^2$ | -0.74 | -0.73 | Wavefront curvature |
| $\dfrac{4\pi\hbar a}{m} n_0 T$ | $2 \times 10^{-4}$ | $-9 \times 10^{-4}$ | Mean field |

**Table 5**: Dominant temperature and density-dependent phase-shifts ($\Delta\phi$) across the Gaussian width of each of the rubidium atomic clouds in QTEST. Mach-Zehnder interferometry is assumed with $T = 10$s, $k_{eff} = 4\pi/780$nm, $R_{eff} = 10^5$ m, $\delta\Omega_y = 10^{-3} \Omega_{ISS}$. $\sigma_r = 6$mm, $\sigma_v \cong 310$ µm/s @ 1nK, and $n_0 = 4 \times 10^5$/cm$^3$ after delta-kick cooling.

*4.6.1 Imaging detection and extraction of the phase difference*

Recent progress in developing high precision atom interferometers have benefited from exploiting phase fringe patterns written on the density-distributions of the AI output ports for measuring and optimizing contrast in a single measurement [9,13], for multi-axis inertial sensing [7], and for quantifying the time evolution of the coherence of Mach-Zehnder AIs with BECs [13]. This development, in conjunction with recent demonstrations of high-stability atom interferometers using simultaneous fluorescence detection with imaging sensors [123,124], leads QTEST to adapt an imaging-based phase detection scheme for the Bragg interferometry, without the internal state labeling or population signal normalization.

After the last π/2 Bragg pulse, the two output clouds from each interferometer separate at a rate of $\hbar k_{\text{eff}}/m \cong 12$ mm/s. This is large compared to the expansion (residual thermal) velocity spread of 0.3 mm/s. After a nominal time of flight, the two clouds will spatially separate. A resonant fluorescence imaging pulse propagating along the interferometer axis will excite the two well-separated clouds simultaneously from each atom interferometer and the scattered photons are collected from the side on an imaging detector, which preserves the spatial features of the clouds [7,9]. Shortly after imaging the outputs of one species ($^{85}$Rb), the same detector records the outputs of the other species ($^{87}$Rb) with another imaging pulse tuned to its resonance.

Ideally, there will be a total of 8 spatial fringe patterns recorded for each measurement, 4 for each atomic species, from which the differential measurement data can be derived. With the principal component analysis (PCA) [125], a well-developed technique adapted by the cold atom community [7,9,10,113,126], the relative phase of $^{85}$Rb vs $^{87}$Rb can be read out at every run. The relative phase is expressed in terms of distance, i.e., the distance between central zero crossings of the two species. It is the sum of the "real" phase difference and imaging delay, where the "real" phase includes the WEP violating phase and all systematics and the imaging delay is calculable with high precision. In actual experiments, one does not need to know the locations of the center zero crossings, but only the relative overall fringe changes between the two species clouds. For a $T$=10 s AI, there will be about 10 fringes across the cloud due to the gravity gradient over the initial cloud size. Assuming an imaging system such that the fringe period on the camera is ~1mm, position error due to camera vibration or timing jitter < 1 micron between two species readout (~1 ms apart) will be sufficient to support AI performance at the atom number shot noise limit of 1000:1. This requirement is easily achievable. Shot-to-shot fluctuation of imaging position error is either statistical (which is then always below shot noise) or has long time scales that will be removed by the various modulation measurements. Here again, QTEST will not try to make an absolute WEP signal determination, but rely on the relative phase versus k-reversal, internal state modulation, and rotating platform modulation to cancel any systematics and reveal the WEP violating phase. Specifically, the WEP signal reverses sign with k-reversal modulation and rotating platform modulation but not with internal state modulation, thus demodulation of the relative phase data at various frequencies allows removal of systematics and bounds the WEP violation.

*4.6.2 Rotation-dependent contrast and its recovery*

The high orbit angular velocity ($\Omega_{\text{ISS}} = 2\pi/91$min) and potentially large dynamic vibrational/rotational environment of ISS have been considered for their effects on the achievable sensitivity of the WEP measurements in QTEST. The dominant mechanism comes from the influence of the Coriolis force on the AIs [127,128], which prevents the interferometers from closing properly and writes a velocity-dependent phase gradient onto the clouds. As shown in Table 5 and Section 5.1, rotations of the Bragg-beam must be suppressed to the level of $10^{-3}$ $\Omega_{ISS}$, corresponding to control of the retro-mirror to at least 1μrad/s to avoid significant rotation-dependent fringes on the clouds. Piezo-actuators on the mirror can achieve this level of precision with sufficient dynamic range, but active feedback from local sensors is required.

Depending on the instrument payload location on ISS and its dynamic environment, two-levels of rotation-stabilization can be envisioned to preserve the AI contrast by compensating the DC and high-frequency noise components of the rotation of ISS with active feedback to piezos on the Bragg retro-mirror. The first level of feedback uses gyros to measure rotational noise and drift in close proximity to the QTEST apparatus to monitor the rotations of ISS at the apparatus. Gyro sensitivities of < 300 nrad/Hz$^{1/2}$ are available in COTS hardware [129]. The second level uses imaging and feedback on the detected AI signal to optimize the contrast for offsets and long-term drifts.

*4.6.3   Gravity-gradient-dependent contrast and its recovery*

The gradient of the gravity field is fundamentally tied to the very concept of WEP tests. The gradient introduces spatial inhomogeneity that results in a large phase shift in a typical Mach-Zenhder atom interferometer under the Earth's gravity field. This leads to contrast loss in each interferometer due to the position and velocity distributions of the atomic cloud. At the same time, since the masses of the two atomic test particles are different, there is no strict common mode cancellation of the AI phase shifts between the two species directly. Precise control in *δr* and *δv* will be able to make the shifts constant and make it possible to cancel in k-reversal measurements.

Consider the dimension along the beam propagation direction. The leading terms of the phase of an open Mach-Zehnder interferometer with pulse spacing of *T* and *T+δT* is:

$$\phi(r_i, v_i) = \phi_0 + k_{\text{eff}}T^2\gamma(r_i + v_iT) + k_{\text{eff}}v_i\delta T, \qquad (9)$$

where $r_i$ and $v_i$ are the initial position and velocity, respectively, $\phi_0 = k_{\text{eff}}gT^2$, and *γ* is the gravity gradient. At an output port, the chance of finding an atom in the diffracted state (excitation probability if state labeled) with such initial conditions is 1-cos$^2\phi(r_i,v_i)$. Assuming an imaging resolution of *σ* and negligible imaging delay after the last interferometer pulse, the observed excitation probability at position $r_f$ with the resolution of *σ* is the average of the excitation probability of all atoms with initial conditions such that their classical trajectories land at position $r_f$ at the detection time: $r_f \approx r_i + 2Tv_i$. Let the initial cloud width and velocity spread be $\sigma_r$ and $\sigma_v$, the observed excitation probability at position $r_f$ can be expressed as:

$$\iint dr_i\, dv_i \exp\left[-\frac{(r_i-\delta r)^2}{2\sigma_r^2}\right]\exp\left[-\frac{(v_i-\delta v)^2}{2\sigma_v^2}\right]\exp\left[-\frac{(r_i+2Tv_i-r_f)^2}{2\sigma^2}\right]\cos^2\phi(r_i,v_i). \quad (10)$$

The integral is proportional to

$$\exp\left[-\frac{(\delta r+2T\delta v-r_f)^2}{2(\sigma^2+\sigma_r^2+4T^2\sigma_v^2)}\right]\left\{1+\exp\left[-\frac{2(k_{\text{eff}}T^2\gamma)^2\left(\sigma^2\sigma_r^2+T^2\left(1-\frac{\delta T}{\gamma T^3}\right)^2\sigma_v^2\sigma_r^2+T^2\left(1+\frac{\delta T}{\gamma T^3}\right)^2\sigma_v^2\sigma^2\right)}{\sigma^2+\sigma_r^2+4T^2\sigma_v^2}\right]\times \quad (11)\right.$$

$$\cos\left[2\left(\phi_0 + r_f \frac{(k_{\text{eff}}T^2\gamma)\left(\sigma_r^2+2T^2\left(1+\frac{\delta T}{\gamma T^3}\right)\sigma_v^2\right)}{\sigma^2+\sigma_r^2+4T^2\sigma_v^2} + \right.\right.$$

$$\left.\left.\frac{(k_{\text{eff}}T^2\gamma)\left(\delta r\left(\sigma^2+2T^2\left(1-\frac{\delta T}{\gamma T^3}\right)\sigma_v^2\right)+T\delta v\left(\left(1+\frac{\delta T}{\gamma T^3}\right)\sigma^2-\left(1-\frac{\delta T}{\gamma T^3}\right)\sigma_r^2\right)\right)}{\sigma^2+\sigma_r^2+4T^2\sigma_v^2}\right)\right]\right\}.$$

The fringe contrast is thus identified as

$$\exp\left[-\frac{2(k_{\text{eff}}T^2\gamma)^2\left(\sigma^2\sigma_r^2+T^2\left(1-\frac{\delta T}{\gamma T^3}\right)^2\sigma_v^2\sigma_r^2+T^2\left(1+\frac{\delta T}{\gamma T^3}\right)^2\sigma_v^2\sigma^2\right)}{\sigma^2+\sigma_r^2+4T^2\sigma_v^2}\right]. \qquad (12)$$

Without imaging detection, where $\sigma \gg \sigma_r + \sigma_v T$, the gravity-gradient dependent phase fringing reduces the contrast, practically limiting $T \leq 5$ s [101]. The contrast can be partially recovered, as proposed in Ref. [108] and also demonstrated experimentally in Ref. [9], by compensating the initial velocity-dependent gravity gradient phase $k_{\text{eff}}T^2\gamma(v_iT)$ with a choice of $\delta T$ so that $1 + \delta T/(\gamma T^3) \ll 1$. However, the remaining contribution from the gravity gradient across the initial cloud size still severely limits the useful range of $T$.

A possible mitigation of the remaining contrast loss comes from imaging detection. By choosing "*anti-compensation*" of the open interferometer asymmetry so that $\beta = 1 - \delta T/(\gamma T^3) \ll 1$, together with the small pixel size approximation $\sigma \ll \sigma_r$ and $\sigma_v T$, contrast is significantly recovered as implied from the above expression of contrast [130]. For instance, assuming $\sigma_r \cong \sigma_v T$ and moderate anti-compensation accuracy, $\beta \leq 10^{-2}$, gravity gradient phase shifts across the initial cloud size of even $k_{\text{eff}}T^2\gamma\sigma_r = 30$ rad correspond to a contrast of >96%.

Now consider directions orthogonal to the beam propagation direction ($z$), as shown in Figure 1 in which the camera points along the $y$ direction. The phase and fringes in the $x$ direction can be treated similar to the $z$ direction considered above. The analogy to the open interferometer asymmetry in the $x$ direction is an additional tilt of the retroreflection mirror in the direction (rotation about y axis) for the last interferometer pulse, which can create spatial fringes in the $x$ direction [7,9]. A parameter similar to $\beta$ can be applied to mitigate contrast loss caused by the kinematics in the $x$ direction with comparable efficiency. The $y$ direction, however, is unique in that there is no imaging resolution in this direction. Since the $y$ direction is chosen to coincide with the rotation vector, leading phase shift terms are much smaller than those from other directions. The impact on contrast is thus negligible.

The analysis of the gravity-gradient-dependent contrast considers only the dimension along $k_{\text{eff}}$. In practice, if $k_{\text{eff}}$ is not aligned to one of the principal axes of the gravity gradient tensor, the fringes will be tilted. On one hand, for the fringes tilted in the imaging plane, detection and extraction of the phase difference would be similar to the one-dimensional case discussed above, assuming the choice of the propagation direction is known when calculating the relative phase. On the other hand, for the fringes tilted perpendicular to the imaging plane, the contrast will be degraded due to phase variations along the imaging line of sight. Such a sinusoidal phase variation of magnitude $p$ would degrade an otherwise perfect contrast to sinc($p$) $\approx 1-p^2/6$, assuming a uniform atomic density distribution. With small misalignment angle between a principal axis of the gravity gradient tensor and the line of sight ($\theta$) of the detector, the magnitude of the phase variation $p$ would be $p \leq 2k_{\text{eff}}\gamma T^2\sigma_r\theta + 4k_{\text{eff}}\sigma_r\delta\Omega T\theta_{\text{rot}}$, where similar effect of 0.1% rotation compensation imperfection ($\theta_{\text{rot}}$) is also included. To maintain 90% (50%) contrast, $\theta \leq$ 13(30) mrad, $\theta_{\text{rot}} \leq$ 172(385) mrad.

*4.6.4 Initial condition sensitivity and mitigation*

The phase terms in the cosine function in the above integral can be grouped into the gravity phase, the gravity gradient phase across the cloud, and the initial condition phase. The gravity phase ($\phi_0$) is the target of the measurement, and the other terms are generally regarded as systematics. However, with imaging detection, the gravity gradient phase across cloud provides a natural and convenient fringe for phase extraction.

The initial condition phase is dependent on the centroids of the clouds ($\delta r$, $\delta v$). Its contribution to the observed excitation probability at position $r_f$, in the limits considered above, reduces to: [130]

$$\frac{(k_{\text{eff}}T^2\gamma)(2T^2\sigma_v^2\delta r - T\sigma_r^2\delta v)}{\sigma_r^2 + 4T^2\sigma_v^2}\beta \approx \frac{(k_{\text{eff}}T^2\gamma)(2\delta r - T\delta v)}{5}\beta. \qquad (13)$$

The sensitivity to the initial conditions is thus suppressed by $\beta$. For this fluctuation not to dominate the shot to shot SNR of 1,000, the stability requirement on dual-species effective CM overlap (taking into account possible distortion of the cloud profiles) reduces to the limits: $\delta r \cong 34$ µm and $\delta v \cong 0.4$ µm/s. This level of systematic CM control should be relatively straightforward to achieve, as discussed in Section 4.5. The gravity direction and k-reversal modulation will then average the WEP signal measurement precision down to the $10^{-16}$ level in a few months of measurements.

## 5 Error Source Analysis

High-sensitivity measurements of gravitational accelerations with AIs require extensive mitigation of systematic shifts to achieve their ultimate accuracy. Rather than trying to achieve the absolute AI measurement precision, QTEST replies on differential acceleration measurements for EP tests, therefore fully taking advantage of common-mode suppression as well as a set of modulated phase sensitive measurements to remove the remaining systematics that are difficult to address directly.

### 5.1 Bragg AI phase shifts and error budgets

Phase calculations for atom interferometers have been well studied for decades. Here we adopt the approach originally developed at Stanford [89], which is generally applicable for non-interacting atoms. In this approach, the classical trajectory of a point particle is calculated for each interferometer arm with a given initial position and velocity. The trajectory is symbolically computed iteratively as a power series of propagation time, which linearizes the influences of Taylor expanded nonlinear potentials while maintaining target accuracy. The photon recoil imparted at each beam splitter is included with magnitude and direction. The classical trajectories are then used to calculate the propagation phase, the laser phase, and the separation phase. The combination of these phases gives the total phase of the interferometer, and is broken down symbolically for physical understanding of the origin of the phase shifts. Table 6 lists the gravity signal and the largest phase shift terms with imperfect rotation compensation and with the assumptions in the table caption.

With atom sources of phase space density of $\sim 10^{-4}$, we consider the mean field shift as additive and will discuss its influence later. Note that the phase terms are not directly measureable as the interferometer readout process averages the sum of all phase terms over initial conditions, as depicted in Section 4.6. The systematic phase shifts of the WEP test are obtained after taking the difference of the corresponding phase terms between two species, as listed in Table 7.

| Phase Term | Phase (rad) | Relative Magnitude | Comment |
|---|---|---|---|
| $k_{\text{eff}} g T^2$ | $-1.40 \times 10^{10}$ | 1 | Reference |
| $k_{\text{eff}} T_{zz} z_0 T^2$ | $2.07 \times 10^3$ | $1.48 \times 10^{-7}$ | Gravity gradient (initial position) |
| $k_{\text{eff}} T_{zz} v_z T^3$ | $4.87 \times 10^2$ | $3.48 \times 10^{-8}$ | Gravity gradient (initial velocity) |
| $k_{\text{eff}}^2 \hbar T_{zz} T^3 / 2m$ | $2.44 \times 10^2$ | $1.74 \times 10^{-8}$ | Gravity gradient due to photon recoil |
| $-2 k_{\text{eff}} v_x \delta\Omega_y T^2$ | $-3.65$ | $2.61 \times 10^{-10}$ | Coriolis |
| $-2 k_{\text{eff}} \Omega_y \delta\Omega_y z_0 T^2$ | $-2.07$ | $1.48 \times 10^{-10}$ | |
| $\frac{7}{6} k_{\text{eff}} T_{zz} \Omega_y v_x T^4$ | $5.47 \times 10^{-1}$ | $3.91 \times 10^{-11}$ | |
| $k_{\text{eff}} B_z \partial_z B \, \hbar m \alpha T^2$ | $4.25 \times 10^{-1}$ | $3.04 \times 10^{-11}$ | Magnetic fields |

**Table 6:** Leading phase terms. For $T=10$ s, initial position $(x_0, y_0, z_0) = (0.01, 0.01, 0.5)$ m, initial velocity $(v_x, v_y, v_z) = (1, 1, 11.8)$ mm/s, ISS rotation $\Omega = (0, \Omega_y, 0) = (0, \Omega_{ISS}, 0)$, residual rotation $\delta\Omega = 10^{-3} \Omega_{ISS} = (1.13, 1.13, 1.13)$ μrad/s. The total potential is expanded as: $-\left(\Phi_0 + g z + \frac{1}{2} T_{zz} z^2 + \frac{1}{3!} Q_{zzz} z^3 + \frac{1}{4!} S_{zzzz} z^4 + \frac{1}{2} T_{xx} x^2 + \frac{1}{2} T_{yy} y^2 + \hbar m \frac{\alpha}{2} (B_z + z\, \partial_z B + x\, \partial_x B + y\, \partial_y B)^2\right)$, where $g = -R_{ISS} \Omega_{ISS}$ at the origin, $B_z = 10$ mG, and $\partial B = 10$ mG/m in all directions. Parameters are evaluated for a $^{87}$Rb F=2, $m_F$=0 interferometer at altitude 400 km above a spherical Earth of radius 6370 km. The pointing of the apparatus is assumed $\theta_x, \theta_y = 1$ mrad off the nadir. A common coefficient $\cos\theta_x \cos\theta_y$ for the terms listed on the table is not shown.

### 5.1.1 Suppression of the dominant phase-shift terms

Suppression and mitigation of each term in Table 7 can be understood intuitively: For terms proportional to $k_{\text{eff}}^2$, which originate from the trajectory difference due to differential photon recoil velocity $\hbar k_{\text{eff}}^2/m$, simultaneous k-reversal measurements will remove all of them because the WEP signal $k_{\text{eff}} g T^2$ changes sign while these terms don't.

The magnetic field gradient dependent shifts in the magnetic insensitive ($m_F = 0$) states are given by the quadratic Zeeman effect, described by the Breit-Rabi formula as $\Delta\omega_{m_F=0} = \pm\frac{1}{2}\alpha B^2$, with $\alpha = (g_J - g_I)^2 \mu_B^2/(h\Delta E_{HFS})$, where $\mu_B$ the Bohr magneton, $g_J$ and $g_I$ are the Landé g-factors, $\Delta E_{HFS}$ is the hyperfine splitting, and the sign corresponds to the atomic hyperfine state ($F = I \pm \frac{1}{2}$) with nuclear spin I. The difference between the isotopes is $\Delta\alpha = 2\pi \times 718.6$ Hz/G$^2$. Earth's field is ~0.5 G and we assume it's shielded to achieve gradients less than 10 mG/m with a 10 mG bias field (longitudinally, to set the quantization axis) inside the science chamber. Because the internal state of the atoms in both interferometer arms are the same, only spatial variations of fields affect the interferometer.

| Phase Term | Phase (rad) | Relative Magnitude | Comment |
|---|---|---|---|
| $\frac{1}{2} k_{\text{eff}}^2 \hbar T_{zz} T^3 \left(\frac{1}{m_{85}} - \frac{1}{m_{87}}\right) c_x c_y$ | $5.7$ | $4.07 \times 10^{-10}$ | Photon recoil |
| $k_{\text{eff}} B_z \partial_z B \, \hbar T^2 \left(\frac{\alpha_{85}}{m_{85}} - \frac{\alpha_{87}}{m_{87}}\right)$ | $5.54 \times 10^{-1}$ | $3.96 \times 10^{-11}$ | Magnetic field |
| $k_{\text{eff}} z_0 (\partial_z B)^2 \hbar T^2 \left(\frac{\alpha_{85}}{m_{85}} - \frac{\alpha_{87}}{m_{87}}\right)$ | $2.77 \times 10^{-1}$ | $1.98 \times 10^{-11}$ | |
| $k_{\text{eff}} v_z (\partial_z B)^2 \hbar T^3 \left(\frac{\alpha_{85}}{m_{85}} - \frac{\alpha_{87}}{m_{87}}\right)$ | $6.52 \times 10^{-2}$ | $4.66 \times 10^{-12}$ | |
| $k_{\text{eff}} T_{zz} \delta v_z T^3$ | $4.14 \times 10^{-2}$ | $2.95 \times 10^{-12}$ | Gravity gradient (initial velocity) |
| $\frac{1}{2 m_{87}} k_{\text{eff}}^2 (\partial_z B)^2 \hbar T^3 \left(\frac{\alpha_{85}}{m_{85}} - \frac{\alpha_{87}}{m_{87}}\right)$ | $3.26 \times 10^{-2}$ | $2.33 \times 10^{-12}$ | Cross term between photon recoil and Zeeman shift |
| $-k_{\text{eff}} B_z \partial_x B \, \Omega_y \hbar T^3 \left(\frac{\alpha_{85}}{m_{85}} - \frac{\alpha_{87}}{m_{87}}\right)$ | $-6.28 \times 10^{-3}$ | $4.48 \times 10^{-13}$ | |
| $k_{\text{eff}} y_0 \partial_y B \, \partial_z B \, \hbar T^2 \left(\frac{\alpha_{85}}{m_{85}} - \frac{\alpha_{87}}{m_{87}}\right)$ | $5.54 \times 10^{-3}$ | $3.96 \times 10^{-13}$ | |

**Table 7:** Leading terms for the phase difference between two species. The magnitude of the phase is calculated assuming initial condition mismatch of 1μm and 1μm/s in all dimensions. $c_x = \cos\theta_x$, $c_y = \cos\theta_y$. A common coefficient $c_x c_y$ to all terms on the table is not shown.

To suppress the B-field dependent systematics, the atom interferometer will be operated subsequently in the two-hyperfine ground state pairs (clock states: $F=1$, $m_F = 0$ and $F=2$, $m_F = 0$ for $^{87}$Rb and $F = 2$, $m_F = 0$ and $F = 3$, $m_F = 0$ for $^{85}$Rb). The sign change of the quadratic Zeeman effect will lead to averaging out of the systematic. However, temporal fluctuations not correlated with the state change cycle lead to stochastic noise. To suppress such noise to below the atom shot noise, the field fluctuations $\Delta(B\delta B)$ need to be less than 0.45 mG$^2$/m. Systematics and drifts can then be removed in data analysis. However, for any B-field dependent systematics correlated with the modulation of the atomic states, the associated component of the field gradient in the interferometer drift tube would have to be suppressed to less than 100 nG/m with ~ 10 mG bias.

After demodulating the data from k-reversal and hyperfine state switching, the remaining terms, given in Table 8, depend on linear combinations of differential initial conditions and gravity gradient or imperfect rotation compensation. Note that the initial position $z_0$ does not appear in the list, which validates the effectiveness of simultaneous k-reversal interferometers using independent dual sources. With tight constraints on differential initial conditions, these phase terms will be below $10^{-6}$ rad. As discussed in Section 4.6.4, imaging of the phase structure and controlled open atom interferometry suppresses the dominant gravity-gradient-dependent systematics. Further, with rotating platform modulation, differential initial conditions will be modulated out while leaving the rest of the terms unchanged. The requirements on initial conditions are thus relaxed, as discussed in Section 3.3.

| Phase Term | Phase (rad) | Relative Magnitude | w/Imaging ($\beta = 10^{-2}$) | w/Rotation Modulation |
|---|---|---|---|---|
| $k_{eff}T_{zz}\delta v_z T^3 c_x c_y$ | $4.14 \times 10^{-2}$ | $2.95 \times 10^{-12}$ | $2.95 \times 10^{-14}$ | $< 3 \times 10^{-18}$ |
| $k_{eff}T_{zz}\delta z_0 T^2 c_x c_y$ | $4.14 \times 10^{-3}$ | $2.95 \times 10^{-13}$ | $2.95 \times 10^{-15}$ | $< 3 \times 10^{-19}$ |
| $-2k_{eff}\delta\Omega_y \delta v_x T^2 c_x c_y$ | $-3.65 \times 10^{-3}$ | $2.61 \times 10^{-13}$ | $2.61 \times 10^{-15}$ | $< 3 \times 10^{-19}$ |
| $\frac{7}{6}k_{eff}T_{zz}\Omega_y \delta v_x T^4 c_x c_y$ | $5.47 \times 10^{-4}$ | $3.91 \times 10^{-14}$ | $3.91 \times 10^{-16}$ | $< 4 \times 10^{-20}$ |
| $-\frac{7}{6}k_{eff}T_{xx}\Omega_y \delta v_x T^4 c_x c_y$ | $2.73 \times 10^{-4}$ | $1.95 \times 10^{-14}$ | $1.95 \times 10^{-16}$ | $< 2 \times 10^{-20}$ |
| $-6k_{eff}\delta\Omega_z\Omega_y \delta v_y T^3 c_x c_y$ | $-1.24 \times 10^{-4}$ | $8.86 \times 10^{-15}$ | | $< 9 \times 10^{-19}$ |
| $-k_{eff}T_{xx}\Omega_y \delta x_0 T^3 c_x c_y$ | $2.34 \times 10^{-5}$ | $1.67 \times 10^{-15}$ | $1.67 \times 10^{-17}$ | $< 2 \times 10^{-21}$ |
| $-k_{eff}T_{xx}\delta v_x T^3 s_x$ | $2.07 \times 10^{-5}$ | $1.48 \times 10^{-15}$ | $1.48 \times 10^{-17}$ | $< 2 \times 10^{-21}$ |
| $k_{eff}T_{yy}\delta v_y T^3 c_x s_y$ | $-2.07 \times 10^{-5}$ | $1.48 \times 10^{-15}$ | | $< 2 \times 10^{-19}$ |
| $-\frac{3}{2}k_{eff}T_{zz}\Omega_y^2 \delta v_z T^5 c_x c_y$ | $-7.97 \times 10^{-6}$ | $5.69 \times 10^{-16}$ | | $< 6 \times 10^{-20}$ |
| $-2k_{eff}\delta\Omega_z\Omega_y\delta y_0 T^2 c_x c_y$ | $-4.14 \times 10^{-6}$ | $2.95 \times 10^{-16}$ | | $< 3 \times 10^{-20}$ |
| $-2k_{eff}\delta\Omega_y\Omega_y\delta z_0 T^2 c_x c_y$ | $-4.14 \times 10^{-6}$ | $2.95 \times 10^{-16}$ | | $< 3 \times 10^{-20}$ |
| $\frac{3}{2}k_{eff}T_{xx}\Omega_y^2 \delta v_z T^5 c_x c_y$ | $-3.98 \times 10^{-6}$ | $2.85 \times 10^{-16}$ | | $< 3 \times 10^{-20}$ |
| $-2k_{eff}\delta\Omega_y \delta v_z T^2 s_x$ | $-3.65 \times 10^{-6}$ | $2.61 \times 10^{-16}$ | | $< 3 \times 10^{-20}$ |
| $\frac{1}{4}k_{eff}T_{zz}^2 \delta v_z T^5 c_x c_y$ | $2.66 \times 10^{-6}$ | $1.9 \times 10^{-16}$ | | $< 2 \times 10^{-20}$ |
| $-k_{eff}T_{xx}\delta x_0 T^2 s_x$ | $2.07 \times 10^{-6}$ | $1.48 \times 10^{-16}$ | | $< 2 \times 10^{-20}$ |
| $k_{eff}T_{yy}\delta y_0 T^2 c_x s_y$ | $-2.07 \times 10^{-6}$ | $1.48 \times 10^{-16}$ | | $< 2 \times 10^{-20}$ |
| $-k_{eff}\delta\Omega_z^2 \delta v_x T^2/\Omega_y\, c_x c_y$ | $-1.83 \times 10^{-6}$ | $1.3 \times 10^{-16}$ | | $< 2 \times 10^{-20}$ |
| $-k_{eff}\delta\Omega_x^2 \delta v_x T^2/\Omega_y\, c_x c_y$ | $-1.83 \times 10^{-6}$ | $1.3 \times 10^{-16}$ | | $< 2 \times 10^{-20}$ |
| $7k_{eff}\delta\Omega_x\Omega_y^2 \delta v_y T^4 c_x c_y$ | $1.64 \times 10^{-6}$ | $1.17 \times 10^{-16}$ | | $< 2 \times 10^{-20}$ |

**Table 8:** Phase terms after k-reversal and internal state modulation, up to $10^{-16}$ relative magnitude. Imaging and control of the phase-fringes of the $^{87}$Rb and $^{85}$Rb clouds further reduces the dominant CM overlap requirements by $\beta$. Modulation of gravity with the rotating platform makes all remaining systematics negligible for measuring $\eta$ to a sensitivity on the order of $10^{-16}$. $s_x = \sin\theta_x$, $s_y = \sin\theta_y$.

Finally, we consider forces from magnetic field gradients external to the rotating AIPP that will limit the usefulness of the rotating platform by imposing CM offsets in position or velocity that are not reversed with respect to gravity by modulation. After transfer to the clock states, to release the atoms from the QUIC trap, stray magnetic field gradients ($\delta B$) will still exert differential accelerations on the $^{87}$Rb and $^{85}$Rb gases, given by the second-order Zeeman shift. One second of free fall time before interferometry, in the bias field and field gradient considered above, leads to an initial center of mass offset in position and velocity of only $\delta r <$ 90 pm and $\delta v <$ 170 pm/s. These effects can be ignored as they are far below any other relevant length scale in the system.

### 5.2 Vibrations

In two-photon Raman interferometers, vibrational noises affect $^{85}$Rb and $^{87}$Rb differently, as detailed in Ref. [100]. The differential sensitivity of vibration comes from different $k_{\text{eff}}$ for $^{85}$Rb and $^{87}$Rb, due to their different hyperfine spacings of ground states in Raman transitions. $k_{\text{eff}}$ can be matched to 2 parts per billion (2 × $10^{-9}$) by choosing laser wavelengths properly. With the QTEST science objective of <$10^{-15}$ g precision performed on ISS, the unsuppressed vibrations are still significant. With the Bragg interferometers proposed for QTEST, $^{85}$Rb and $^{87}$Rb can be driven by identical frequency components with matching diffraction efficiency, due to the accessibility of the "magic" wavelength. Thus, $k_{\text{eff}}$ and the momentum transferred to the atoms are identical, directly translating to identical sensitivities to accelerations. The vibrational noise induced phase shifts are then strictly common to the two species. Detunings deviating from the magic wavelength will affect the relative diffraction efficiency, while $k_{\text{eff}}$ is always perfectly matched. The impact on contrast will be second order to diffraction efficiency variation when the Mach-Zehnder interferometer operates at the optimal condition. The requirement on environmental vibration is thus relaxed to the Doppler shift when the atom interferometer light is on and bounces off of moving optics. Let the allowance of Doppler shift noise be 400 Hz (assumed to be 10% of the two-photon Rabi frequency), corresponding to an RMS velocity of 157 μm/s. For comparison, a vibrational noise of $10^{-4}$ m/s²/√Hz would have RMS velocity noise of 71 μm/s, integrating from 0.05 Hz to infinity.

### 5.3 Mean field shifts

Mean field systematic shifts arise from inter- and intra-particle collisions among the atoms in ultracold, dual-species atomic gases. Here, only s-wave scattering is generally allowed, and the interactions are fully described by the s-wave scattering length. For non-degenerate, dual-species $^{85}$Rb-$^{87}$Rb gases, the mean-field energy shift per particle is given by:

$$\hbar\omega_{87} = \frac{4\pi\hbar^2}{m_{87}}(a_{87-87}n_{87} + a_{85-87}n_{85}), \quad (14)$$

$$\hbar\omega_{85} = \frac{4\pi\hbar^2}{m_{85}}(a_{85-85}n_{85} + a_{85-87}n_{87}), \quad (15)$$

where $n_x$ is the average density of the rubidium atomic gas (species x), $m_x$ is the atomic mass, and $a_{x-y}$ are the interspecies and intraspecies scattering lengths. In the atom interferometer, imperfect beam-splitting efficiency of the first π/2 pulse leads to a phase-shift: $\Delta\phi_x = \int_0^{2T} \omega_x \chi_x \, dt$, where $T=10$ s is the interferometer time and $\chi_x$ is the fractional imbalance of atomic populations in the two interferometer arms. The table below, characterizing the magnitude of the interaction shift for the QTEST dual-species gases assuming $\chi_x = 10^{-2}$, integrates the time-dependent density to account for ballistic expansion but doesn't include corrections due to the separation of the two species' CM during interferometry (inter-species scattering). Note that the mean-field effect is very small without additional suppression. However, since the phase term is independent of $k_{eff}$, the mean-field effect will be modulated further with k-reversal (limited only by our knowledge of the average atomic gas density at each trap).

|  | $a_{bg}$ | Phase (rad) | Relative Magnitude |
|---|---|---|---|
| $^{87}$Rb | 100 $a_0$ | $2 \times 10^{-6}$ | $1.2 \times 10^{-16}$ |
| $^{85}$Rb | -450 $a_0$ | $-9.1 \times 10^{-6}$ | $-5.8 \times 10^{-16}$ |
| $^{87}$Rb-$^{85}$Rb | 213 $a_0$ | $4.2 \times 10^{-6}$ | $2.7 \times 10^{-16}$ |

**Table 9:** Interspecies and intraspecies mean-field parameters for the low-temperature, nondegenerate $^{85}$Rb-$^{87}$Rb mixture. The magnitudes of the phase shifts are further suppressed to be negligible with k-reversal.

### 5.4 Wavefront related effects

The atomic phase during interferometry is greatly affected by fluctuations of the Bragg-beam collimation and motion of the atoms across the effective Bragg-laser wavefront. Such motion can induce sizable systematic shifts and statistical fluctuations attributed to insufficient collimation and purity of the incident and reflected lasers [106,131,132], and can also write phase gratings onto the atomic clouds to reduce the contrast (notably in a direction perpendicular to the imaging plane). To first order, the wavefront-curvature-dependent phase shift is given by:

$$\Delta\phi_{WF}(z,t) = -\frac{1}{2}\frac{k_{eff}}{R_{eff}(z)}(\sigma_r + \sigma_v t)^2, \qquad (16)$$

where $R_{eff}(z) = (1/R_I(z) + 1/R_R(z))^{-1}$ is the sum of the incident ($R_I$) and reflected ($R_R$) radii of curvature of the Gaussian beam at a distance (z) from the waist and $t$ is the free expansion time. In the ideal case, the 24 mm waist beam is collimated at the retro-mirror, such that $R_{eff}(z) \cong z[1+(z_R/z)]^2 > 5 \times 10^6$ m over a $z = 1$ m beam path-length. Here, the thermal velocity for $^{87}$Rb and $^{85}$Rb atoms at 1 nK ($\sigma_v \cong 310$ μm/s), and assuming initial cloud widths $\sigma_r = 6$ mm, leads to a phase-shift across the cloud of approximately $6 \times 10^{-5}$ radians, negligibly effecting the contrast.

A more reasonable approximation for the effective Bragg wavefront curvature considers the effective wavefront dependence on the initial collimation [133]. For a fiber-based collimator producing a 24 mm waist beam (fiber numerical aperture = 0.12, focal length = 0.2 m, λ = 780 nm), a longitudinal displacement of the fiber-lens spacing of even 100 μm still allows $R_{eff}(z) > 10^5$ m over the ~0.5 m interaction region. The resultant phase shift on each rubidium AI is calculated as:

$$\Delta\phi_{WF}(z_0, t_0) - 2\Delta\phi_{WF}(z_1, t_0 + T) + \Delta\phi_{WF}(z_2, t_0 + 2T), \qquad (17)$$

where $z_0$, $z_1$, and $z_2$, are the cloud CM positions at the first, second, and third interferometer pulses, respectively. The level of collimation assumed leads to a systematic error $\Delta\eta_{WF} \cong 8 \times 10^{-14}$. Wave-front curvature at this level will moderately affect the contrast (see Table 5), random fluctuations of the fiber-lens spacing of even ±10 μm RMS will be below shot-noise by over an order of magnitude, and systematic offsets attributed to differential CM of the clouds and differential cloud widths/temperatures will be removed by k-reversal and demodulation with the rotation platform. After demodulation, the wavefront related shifts will be suppressed to $\Delta\eta_{WF} < 10^{-17}$.

### 5.5 Bragg laser noise and the diffraction phase

Laser frequency noise introduces a relative phase shift between the incident and time-delayed return components constituting the Bragg lattice. The two rubidium species are identically affected by the laser phase noise, suppressing its influence for differential acceleration measurements. However, CM separations of the gases during interferometry give, in the case of white frequency noise, a residual sensitivity of [134]:

$$\sigma_\phi^2 \approx \frac{4\pi^3}{\tau}\frac{L^2}{c^2}\Gamma. \qquad (18)$$

where $L$ is the mean displacement of the atomic clouds of each species along the laser k-vector, $c$ is the speed of light in vacuum, and $2\tau = \pi/\Omega$ with $\Omega$ the two-photon Rabi frequency. Assuming that the Bragg laser has a line width $\Gamma < 1$ MHz and $\tau > 10\mu s$, gives $\sigma_\phi < 10^{-4}$ rad, which is much smaller than the atomic shot noise.

Interferometers using Bragg diffraction have been known to exhibit diffraction phases [22,94,127]. In light-pulse AIs, these systematic shifts arise from operating in the quasi-Bragg regime, where atom-light interactions cause atom losses from the effective two-level system and thereby break down the AI time-reversal symmetry: In absence of such losses, the second half of the AI is a time-reversed mirror image of the first, and any diffraction phases cancel, unless the symmetry is broken by technical imperfections. The presence of coherent loss channels breaks this symmetry. A model of these effects can be obtained by numerically integrating the Schrödinger equation for the atom-light interaction, and has been shown to agree with experimental data. For $\hbar k_{eff}$ Bragg diffraction in the short-pulse regime used in QTEST, the uncompensated diffraction phase per beam splitter can amount to hundreds of milliradians [95].

In the QTEST implementation, several factors will reduce the diffraction phase to a level below the science requirement. First, Mach-Zehnder interferometers are inherently free of diffraction phases if the atoms are detected at the output that moves with the same momentum as the input (detecting the other output leads to the same results only if there are no losses). QTEST will detect the fringes from both ports for each interferometer. In addition, QTEST will use much longer pulse separation times than used in [95], thereby reducing the relative influence of the shift, and longer pulse durations, therefore strongly suppressing the diffraction phases. According to the diffraction phase model, simultaneously using two species having the same $\Omega$ will mostly have a common diffraction phase. Residual diffraction phase shifts will be suppressed with k-reversal. All remaining systematics are independent of the influence of gravity (e.g. Doppler dependent terms) and therefore will be averaged to zero with the rotating platform modulation measurements.

### 6 Conclusions

We have investigated QTEST, an ISS mission concept that performs high precision atom interferometer experiments in the microgravity environment in space. The measurements with quantum test masses will probe the nature of gravity and quantum mechanics, and explore dark matter and quantum electrodynamics. The primary goal of QTEST is to perform WEP measurements with two isotopes of rubidium ($^{87}$Rb and $^{85}$Rb) to a precision better than 10$^{-15}$, not only improving over the best classical tests by two orders of magnitude, but also extending the precision of quantum WEP tests by a factor of over $10^6$. By extending the WEP validity range with rare isotopes and quantum test masses, it provides access to spin-gravity coupling effects at the same time, and therefore is complementary to the WEP measurements with classical test masses. The high performance measurement capabilities of atom interferometers in QTEST will also be able to support a number of other precision measurements. QTEST defines its secondary goals for photon recoil measurements, to better than $10^{-11}$ precision, enabling highly accurate tests of the theory of quantum electrodynamics and demonstrating the feasibility of establishing $^{87}$Rb and $^{85}$Rb as calibrated microscopic mass standards for use worldwide.

QTEST will fully exploit the microgravity environment on ISS to achieve long free fall times in a well-controlled environment, and resultant measurement sensitivities that are orders of magnitude higher than those accessible on earth. The space environment enables QTEST to modulate the gravity signal by rotating the apparatus on a rotating platform to remove otherwise difficult systematics associated with, *e.g.*, the system design and initial sample conditions. QTEST incorporates a combination of systematic reduction approaches (rotating platform modulation, simultaneous $k_{\text{eff}}$ reversal with two source regions, and hyperfine state modulation), and techniques at the leading edge of AI technologies (Bragg interrogation at the "magic" wavelength for $^{85}$Rb and $^{87}$Rb, imaging-based phase-detection, delta-kick cooling to achieve ultra-low temperatures, etc.). These innovations make QTEST unique, enabling fundamental science at unprecedented precision in the relatively noisy environment on the ISS platform.

Finally, QTEST will leverage NASA's Cold Atom Lab (CAL), a mission onboard ISS planned for flight in 2017, as a technology demonstrator [135]. The CAL mission includes several critical milestones on the way to atom interferometry in space. Among these are the first ultra-cold atoms in space under microgravity, achieving 100 pK ultra-cold atom ensembles, long free expansion time (> 5 seconds) in microgravity, Bragg beam splitting/atom transport, laser operation for cold atom generation and control, as well as ultra-high vacuum (UHV) technology on ISS. CAL as a multi-user facility will have principle investigator-led investigations with overlapped dual-species ($^{39}$K-$^{87}$Rb) interferometry with a single Bragg laser at the "magic" wavelength of 785nm that demonstrates initial space-based measurements for WEP and photon recoil measurements. The QTEST design and implementation will be built on the CAL heritage.

**Acknowledgements**

The authors wish to acknowledge useful discussions with Rob Thompson, Ernst Rasel, Markus Krutzik, Justin Khoury, Jay Tasson, and Eric Copenhaver. We are especially appreciative of assistance from Surjeet Rajendran for contributing the ultralight dark matter study. We also want to thank Jason Hogan and David Johnson for use of their software package, which was originally developed and validated by the Kasevich team at Stanford University, and was modified for QTEST. Government sponsorship is acknowledged. This work was carried out at the Jet Propulsion Laboratory, California Institute of Technology, under a contract with the National Aeronautics and Space Administration (NASA). H.M. acknowledges support from NASA. Copyright 2015. All rights reserved.